\documentclass[tighten,nofootinbib,aps,amssymb,floatfix]{revtex4}
\newcommand{\be}{\begin{equation}}
\newcommand{\ee}{\end{equation}}
\newcommand{\beq}{\begin{eqnarray}}
\newcommand{\eeq}{\end{eqnarray}}

\usepackage{bm}
\usepackage{graphicx}%
\usepackage{epsf}
\usepackage{epsfig}

 \usepackage{amsmath}
\usepackage{amsfonts,amssymb}

\usepackage{dsfont}
\usepackage{slashed}
\usepackage{booktabs}

\graphicspath{{.}{Graphics/}}                    

\newcommand{\C}{\mathcal{C}} 
\newcommand{\eins}{\mathds{1}} 

\begin{document}


\title{Evaluation of fermion loops applied to the
calculation of the $\eta^\prime$ mass and the nucleon scalar and electromagnetic form factors}

\author{C.~Alexandrou~$^{a, b}$, K. Hadjiyiannakou~$^{a, b}$, G. Koutsou~$^b$
 A. \'O Cais~$^b$ and A. Strelchenko~$^b$}
\affiliation{$^a$ {Department of Physics, University
  of Cyprus, PoB 20537, 1678 Nicosia, Cyprus \\
  $^b$  Computation-based Science and
  Technology Research Center, The Cyprus Institute, 
P.O. Box 27456, 1645 Nicosia, Cyprus}}
\email{alexand@cyi.ac.cy (C.Alexandrou),hadjigiannakou.kyriakos@ucy.ac.cy (K.Hadjiyiannakou) ,g.koutsou@cyi.ac.cy (G. Koutsou), a.strelchenko@cyi.ac.cy (A. Strelchenko)}

\date{\today}
       
\begin{abstract}
 
The exact evaluation of the disconnected diagram contributions 
to  the flavor-singlet pseudo-scalar meson mass, the 
nucleon $\sigma$-term and the nucleon electromagnetic form factors is carried out utilizing GPGPU technology 
with  the NVIDIA CUDA platform.  
The disconnected loops 
are also computed using stochastic methods  with several noise reduction techniques.
 Various dilution schemes as well as the  truncated solver method are studied. We make a comparison of these stochastic techniques to the exact results and show that the number of noise vectors depends on the operator insertion in the 
fermion loop.
\end{abstract}

\keywords{Lattice QCD, nucleon form factors, nucleon sigma term }

\maketitle

\section{Introduction}
An accurate estimate of disconnected contributions to flavor singlet quantities remains one of the most computational demanding problems
in applying lattice Quantum Chromodynamics (QCD) to the study of hadron 
physics.  The determination, for instance of the strange content of the nucleon,
which has been extensively studied both experimentally~\cite{Aniol:2005zf,Armstrong:2005hs,Maas:2004dh,Maas:2004ta,Spayde:2003nr} and 
theoretically, requires the evaluation of fermion loops.
 A nice example of an analysis that combines lattice results and
chiral expansion techniques to determine  the strange magnetic and electric
form factors of the nucleon is presented in Refs.~\cite{Leinweber:2004tc,Leinweber:2006ug}.
The exact evaluation of disconnected diagrams 
is extremely difficult because one needs 
to calculate the all-to-all propagators. 
Furthermore, the gauge noise for some of these disconnected diagrams dominates the signal
and a large  number of statistics is required to reduce the error.
To avoid performing all the inversions required for an exact evaluation of the all-to-all propagators, the standard approach is to use stochastic techniques with
a variety of dilution schemes to estimate them. Such techniques have been applied recently in the evaluation of the $\eta^\prime$ mass, the nucleon $\sigma$-term and the electromagnetic form factors 
 and the hadronic contribution to $g-2$~\cite{Jansen:2008wv,Bali:2009hu,Doi:2009sq,Babich:2010at,Takeda:2010cw,Feng:2011zk}.
The aim of this work is to evaluate a representative
set of disconnected loops {\it exactly} and compare to the routinely used 
stochastic techniques in order to benchmark the various approaches.
Recently one  utilizes special hardware accelerators, 
such as graphics processors, 
to speed-up the inversions themselves~\cite{Barros:2008rd,Alexandrou:2010jr}.
The exact evaluation is thus carried out  using Graphics Cards (GPUs)
 to efficiently calculate 
 the all-to-all propagator.
Since the purpose of this work is to benchmark the various methods
rather than produce state-of-the-art results, we use
 gauge configurations generated by the SESAM Collaboration~\cite{Hoeber:1997rg,Orth:2005kq} on 
a relatively small lattice size of $16^3 \; \times \; 32$ in order
to facilitate the exact evaluation of the fermion loops.
We examine various fermion loops that enter into the evaluation of observables
that have been recently studied on the lattice. These are the mass of
the $\eta^\prime$ and the nucleon scalar and electromagnetic (EM) form factors.

\section{Lattice ensemble and simulation parameters}
For this exploratory study we use  gauge configurations generated by the SESAM/T$\chi$L Collaboration using  $N_f=2$
 Wilson fermions
at $\beta=5.6$  and hopping parameter $\kappa=0.157$. This
corresponds to a pion mass of $am_\pi=0.3452(29)$~\cite{Orth:2005kq}. 
In order to convert to physical units we  follow Ref.~\cite{Orth:2005kq}
and use the Sommer parameter, $r_0$, defined through the force between 
two static quarks at intermediate distance~\cite{Sommer:1993ce}.
 The value taken from  Ref.~\cite{Orth:2005kq} is $r_0=0.5$~fm
and at $\kappa=0.157$   it yields for  the
inverse lattice spacing  $a^{-1}=2.16(3)$~GeV, giving 
$m_\pi=746$~MeV. A total of 165 configurations are analyzed.

  The general form of a  disconnected loop is given by
\begin{equation}
L(x)=Tr[\Gamma G(x;x)],
\label{loop}
\end{equation}
where for  $\Gamma$ we consider $\Gamma={\bf 1}, \,\gamma_5,$ and $\gamma_\mu$ and
$G(x;x)$ is the 
 Dirac propagator. 
The main question that we would like to address is which type of stochastic 
technique 
 most efficiently reproduces the exact result and whether an optimal method  exists independently of the $\Gamma$ insertion in the loop.
It is clearly seen from the form of Eq.~(\ref{loop}) that, in order to evaluate quantities that involve the  spatial sum of $L(x)$,  one needs spatial volume more inversions than the point-to-all propagator. In this work,
we evaluate exactly the all-to-all propagators for the particular set of parameters, given above. This is clearly very computational intensive  
 and it is therefore beneficial to take advantage of graphics accelerators.
 We employ the QUDA library, which
  provides mixed precision implementations of conjugate gradient (CG) and BiCGstab solvers for 
the NVIDIA CUDA platform, in order to evaluate the all-to-all propagators~\cite{Clark:2009wm}. The exact result then provides a benchmark at the level of gauge-noise for the quantities with contributions from disconnected loops. We note that GPUs are used for performing  all the inversions including those for the stochastic evaluation.  We would like
to point out that increasing the number of configurations to reduce the 
gauge noise is expensive since this would require 
evaluating more all-to-all propagators. Therefore in this work we limit
ourselves to 165 gauge configurations.

The stochastic estimate of the disconnected  quark loops is performed using 
complex  $Z_2$ noise for the source vectors in combination  with several dilution schemes and the truncated solver method~\cite{Bali:2009hu}. Specifically, 
 we consider space, color and spin dilution schemes. 
Color (spin) dilution requires  three (four) times more inversions as compared to the number with
no dilution, whereas even-odd partitioning of the space
doubles that number.
 In addition to an even-odd dilution, we have also applied a \emph{cubic} dilution, where separate sources are placed on each vertex of an elementary 3-dimensional cube and repeated throughout the lattice, leading to an increase of a factor of 8 in the number of inversions. 
The truncated solver method effectively partitions the problem into
 a low precision and high precision space~\cite{Bali:2009hu}. 
A large number
of low precision inversions are carried out to achieve 
an approximation to the propagator with low stochastic error (but only accurate to low precision). A high precision stochastic correction 
is then applied using a small ensemble with the corresponding inversions carried out to high precision.  The size of the stochastic ensemble of  noise vectors 
for the low precision space and the corresponding  ensemble of noise vectors for the high-precision correction is examined for the various loops.
Time dilution
is applied in all cases and we exploit translational invariance in order to limit the number of time slices for which  the exact evaluation of the fermion loops
is required.

The  stochastic evaluation schemes
are  based on creating an ensemble of
noise-vectors with the properties
\begin{equation}
\frac{1}{N_r}\sum_{r=1}^{N_r} \xi_\mu^a(x)_r\; = \; \langle \; \xi_\mu^a(x) \; \rangle_r \; \approx \; 0
\end{equation}
and
\begin{equation}
\frac{1}{N_r}\sum_{r=1}^{N_r} \xi_\mu^a(x)_r \xi_{\mu^\prime}^{* a^\prime}(x^\prime)_r \; = \; \langle \; \xi_\mu^a(x) \xi_{\mu^\prime}^{* a^\prime}(x^\prime) \; \rangle_r \; = \; \delta(x-x^\prime) \delta_{\mu\mu^\prime} \delta_{a a^\prime}\quad .
\label{eq: delta functions}
\end{equation}
Using the above properties one has
\beq
\langle \; \phi_\mu^a (x) \xi_\nu^{*b}(y) \; \rangle_r \; &= &\; \sum_{y^\prime} G_{\mu \nu^\prime}^{a b^\prime} (x;y^\prime) \langle \; \xi_{\nu^\prime}^{b^\prime}(y^\prime) \xi_\nu^{* b} (y) \; \rangle_r
\nonumber \\
&=&\sum_{y^\prime} G_{\mu \nu^\prime}^{a b^\prime} (x;y^\prime) \delta(y-y^\prime) \delta_{\nu\nu^\prime} \delta_{b b^\prime}
=G_{\mu\nu}^{ab}(x;y)\quad,
\label{eq: propagator noise*result}
\eeq
where $\phi$ is the solution vector corresponding to a source with
 noise vector $\xi$. 
The above equation provides an approximation to the exact 
all-to-all propagator because the property of the noise vectors $\xi$ 
given in Eq.~(\ref{eq: delta functions}) is exact only for
 an infinite ensemble of vectors.
 In practice, one takes a large number of noise vectors and tests the stability of results when increasing the number of vectors.
In this work, we will determine how large the number of noise vectors
 should be, in order to obtain the exact result.

Using the stochastic estimate of the all-to-all propagator the  disconnected loop is written as
\begin{equation}
L(\vec{x},t_0)=\frac{1}{N_r} \sum_{r} \xi_\alpha^{*c}(\vec{x},t_0)_r \Gamma_{\alpha \beta} \phi_\beta^c(\vec{x},t_0)_r.
\label{eq: loop stochastic}
\end{equation}
Given that Eq.~(\ref{eq: loop stochastic}) provides
 an estimate to the exact result the question is whether the
   size of the noise  vector ensemble depends on the type of $\Gamma$ matrix
in the loop.

In the following sections we discuss the results according to the different 
choices of the $\Gamma$-matrices entering in the loop.

\section{Evaluation of the $\eta^\prime$ mass}

A general non-flavor singlet meson correlator $C_{AB}({\bf p},t)$ can be written as
\beq
C_{AB} ({\bf p},t_f-t_i)& = &  \frac{1}{L^3}\sum_{{\bf x}_f,{\bf x}_i}\langle J_A({\bf x_f},t_f)J_B^\dagger({\bf x}_i,t_i) \rangle \nonumber \\
&=&\frac{1}{L^3}\sum_{{\bf x}_f,{\bf x}_i}\langle e^{i{\bf p}.({\bf x}_f-{\bf x}_i)}\textrm{Tr}\left [\Gamma_A G_{2}({\bf x}_f,t_f;{\bf x}_i,t_i)\Gamma_B G_{1}({\bf x}_i,t_i;{\bf x}_f,t_f)\right]\rangle,
\label{eqn:all_corr}
\eeq
where the interpolating field $J_A(x)=\bar{q}_{1}\Gamma_A q_{2}$ for different quark flavors $q_1$ and $q_2$, and  $G_{j}({\bf x^\prime},t^\prime;{\bf x},t)$ is the propagator of quark of flavor $j$  from space-time point $({\bf x},t)$ to space-time point $({\bf x^\prime},t^\prime)$.  Spin and color indices are suppressed.
For flavor singlet mesons such as the $\eta^\prime$,  besides the connected
part   given by Eq.~\ref{eqn:all_corr}, there are
disconnected contributions to the correlation function, which are given by
\begin{equation}
D_{AB}({\bf p}, t_f-t_i) =  \frac{1}{L^3}\sum_{{\bf x}_f,{\bf x}_i}\langle \textrm{Tr}\left[e^{i{\bf p}.{\bf x}_f}\Gamma_A G_{2}({\bf x}_f,t_f;{\bf x}_f,t_f)\right] {\rm Tr}\left[e^{-i{\bf p}.{\bf x}_i}\Gamma_B G_{1}({\bf x}_i,t_i;{\bf x}_i,t_i)\right]
 \rangle.
\label{eqn:all_dis_corr}
\end{equation}

Smearing is routinely used to 
decrease overlap with excited states. 
In this work, in addition to local,  we consider Gaussian smeared quark  
fields~\cite{Alexandrou:1992ti,Gusken:1989} for the construction of  
the interpolating fields:  
\beq  
q_{\rm smear}^a(t,{\bf x}) &=& \sum_{\bf y} F^{ab}({\bf x},{\bf y};U(t))\ q^b(t,{\bf y})\,,\\  
F &=& (\eins + {\alpha} H)^{n} \,, \nonumber\\  
H({\bf x},{\bf y}; U(t)) &=& \sum_{i=1}^3[U_i(x) \delta_{x,y-\hat\imath} + U_i^\dagger(x-\hat\imath) \delta_{x,y+\hat\imath}]\,. \nonumber  
\eeq  
In addition, we apply APE-smearing to the gauge fields $U_\mu$ entering   
the hopping matrix $H$.  
All forward point-to-all propagators are computed by applying smearing  
taking the values of the smearing parameters $\alpha=4.0$ and $n=50$.
These values were determined by  optimizing 
ground state dominance  for the nucleon~\cite{Alexandrou:2008tn}. 
The loops are computed without smearing
throughout, including those involved in the computation of the $\eta^\prime$ mass. 
Since the purpose of this work is to compare exact results with
those using various stochastic approaches we did not repeat the
evaluation of the exact loops with smearing.

For the particular case of the $\eta^\prime$ meson and since we are using 
 an $N_f=2$ gauge ensemble, there  are no strange quark contributions. Therefore   the flavor singlet pseudo-scalar meson (also denoted by $\eta_2$ in the $N_f=2$ theory) has only contributions from the light quarks
and   its two-point correlator can be written as
\begin{equation}
C_{\eta^\prime}(t) = C_{\pi}(t) - 2D(t),
\end{equation}
where we have taken ${\bf p}=0$  and
dropped the flavor
indices $f_1$ and $f_2$. 

\begin{figure}[h!]
\begin{minipage}{0.49\linewidth}
{\includegraphics[angle=-90,width=\linewidth]{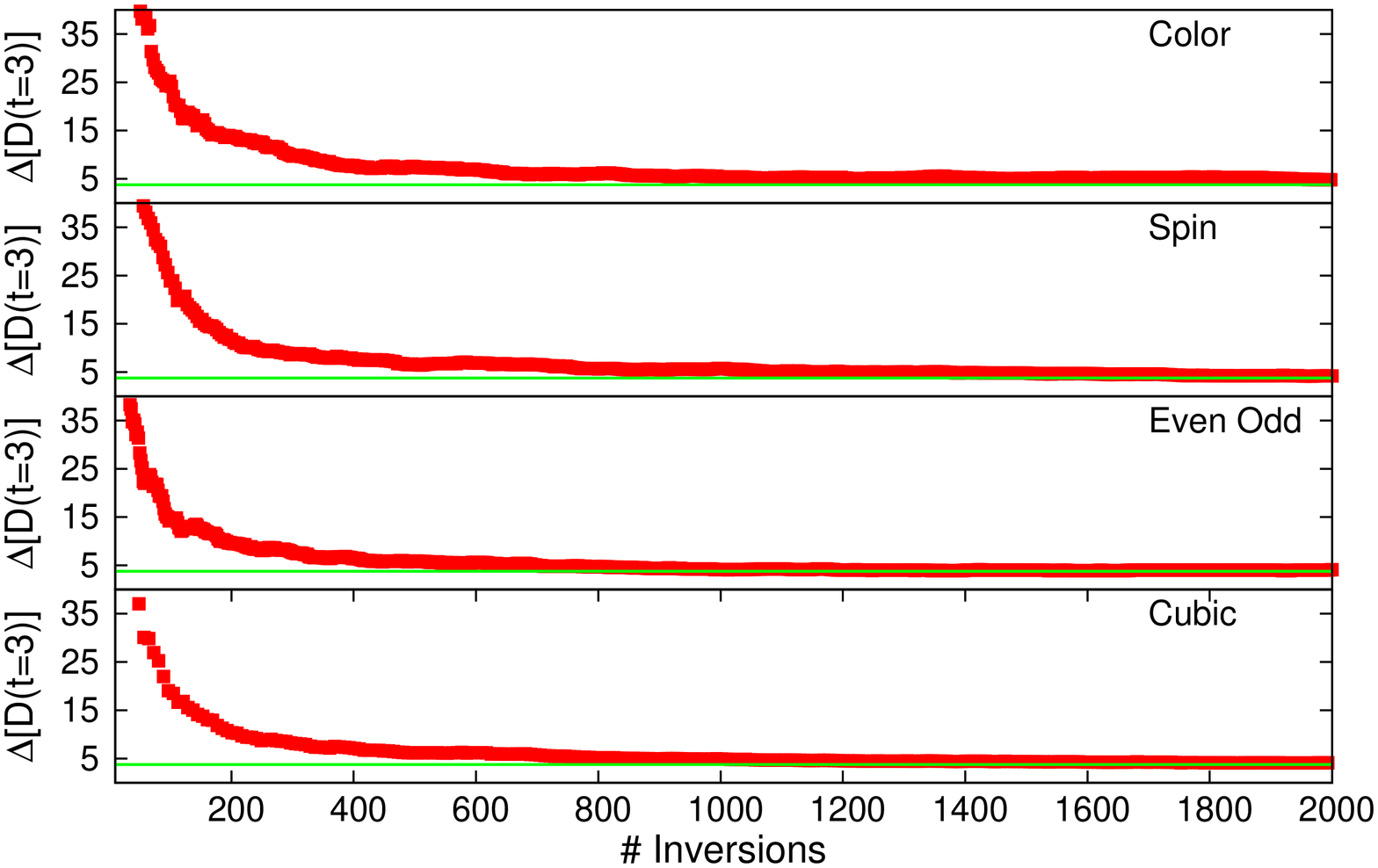}}
\end{minipage}
\begin{minipage}{0.49\linewidth}
{\includegraphics[angle=-90,width=\linewidth]{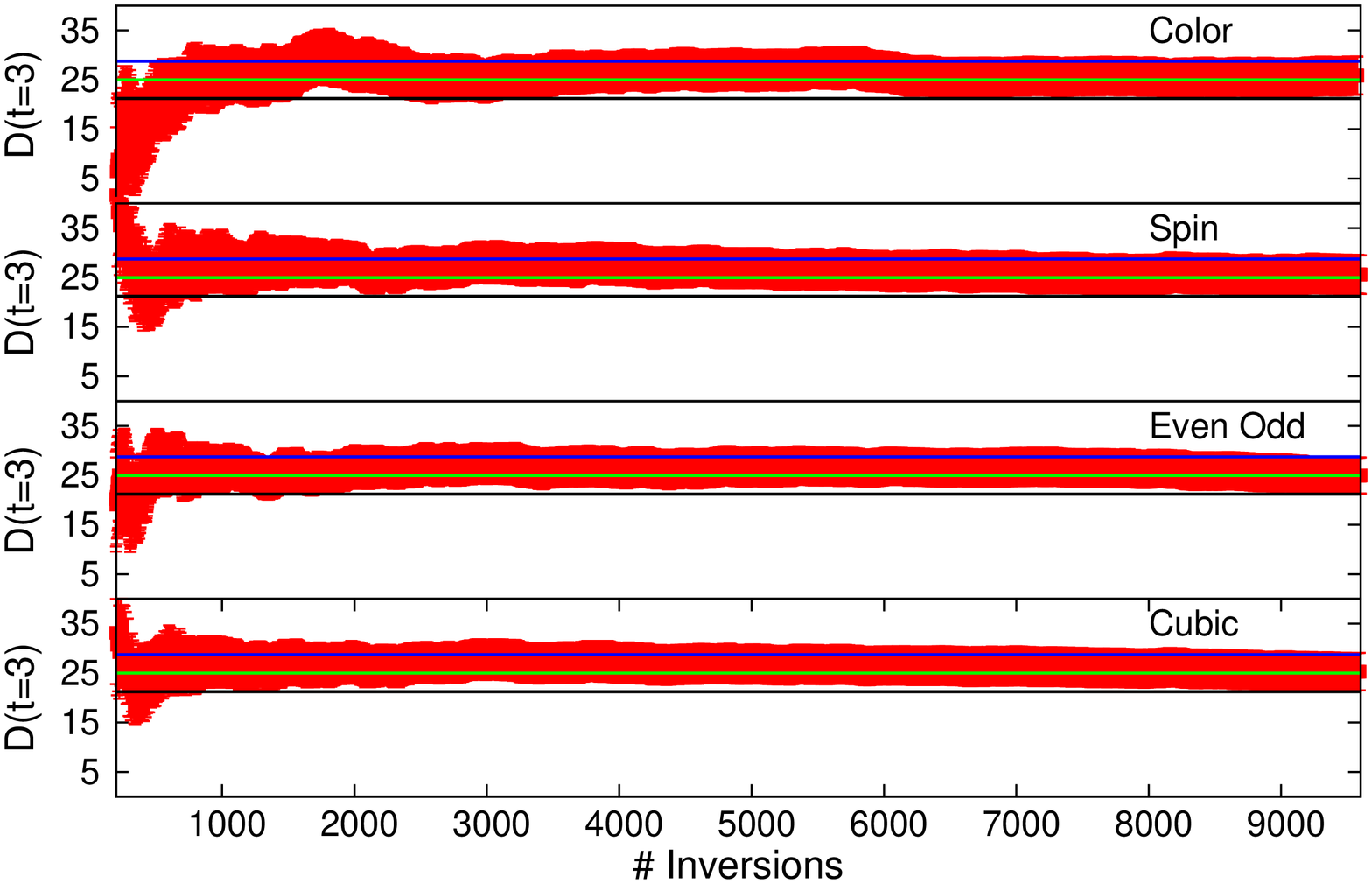}}
\end{minipage}
\caption{Left panel: The error on the  disconnected part of the $\eta^\prime$ correlator $D(t)$ at $t/a=3$ as a function of the number of inversions. The line 
shows the statistical error. Right panel: The   disconnected part of the $\eta^\prime$ correlator $D(t)$ at $t/a=3$ computed stochastically
as a function of the number of inversions.  The lines show the mean value and error band of the exact result for
$D(t)$ at the same time slide.
 In both graphs we show, from top to bottom, results using:  color, spin, even-odd and cubic
dilution.} 
\label{fig:compare dilution}
\end{figure}

\begin{figure}[h!]
\begin{minipage}{0.49\linewidth}
{\includegraphics[angle=-90,width=\linewidth]{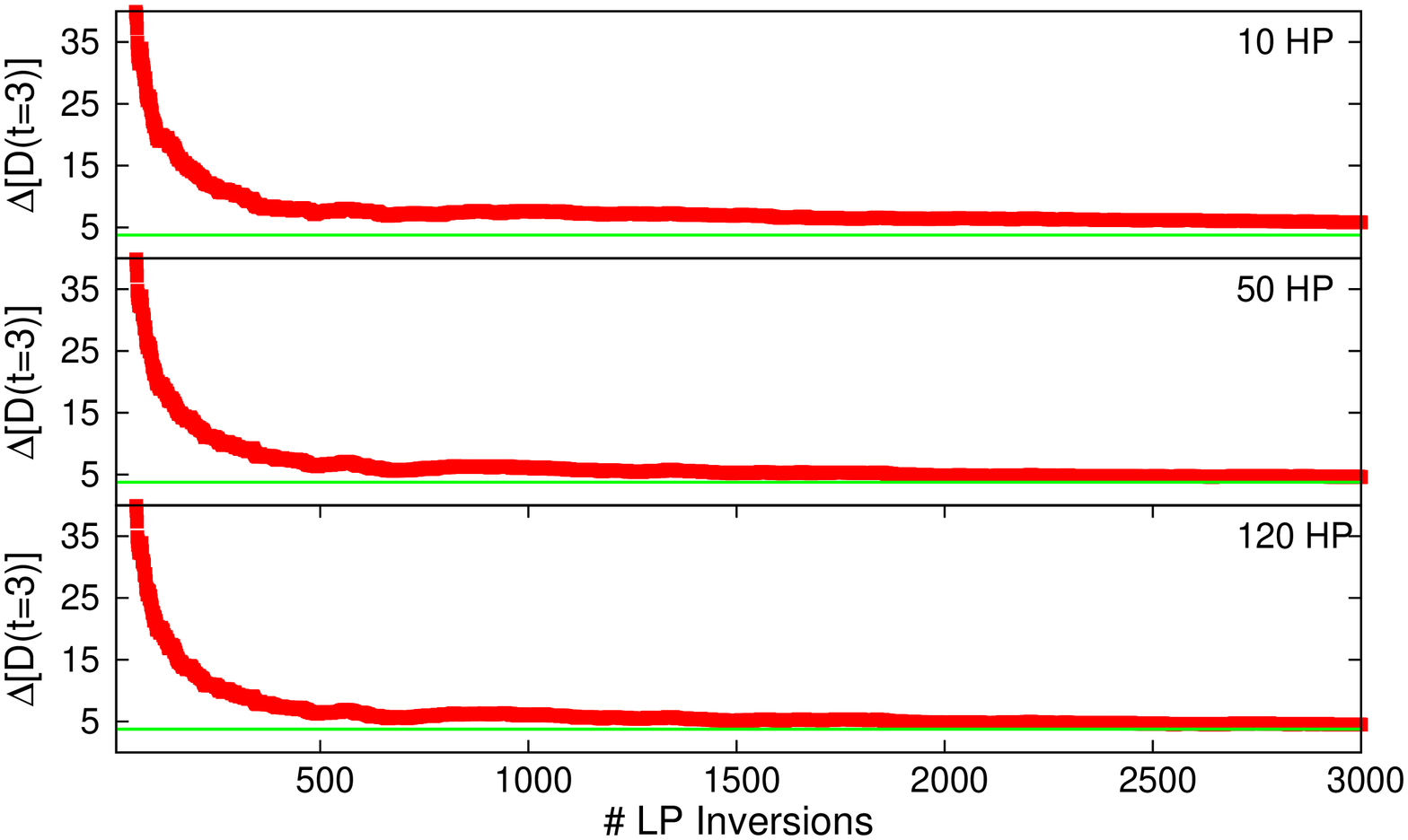}}
\end{minipage}
\begin{minipage}{0.49\linewidth}
{\includegraphics[angle=-90, width=\linewidth]{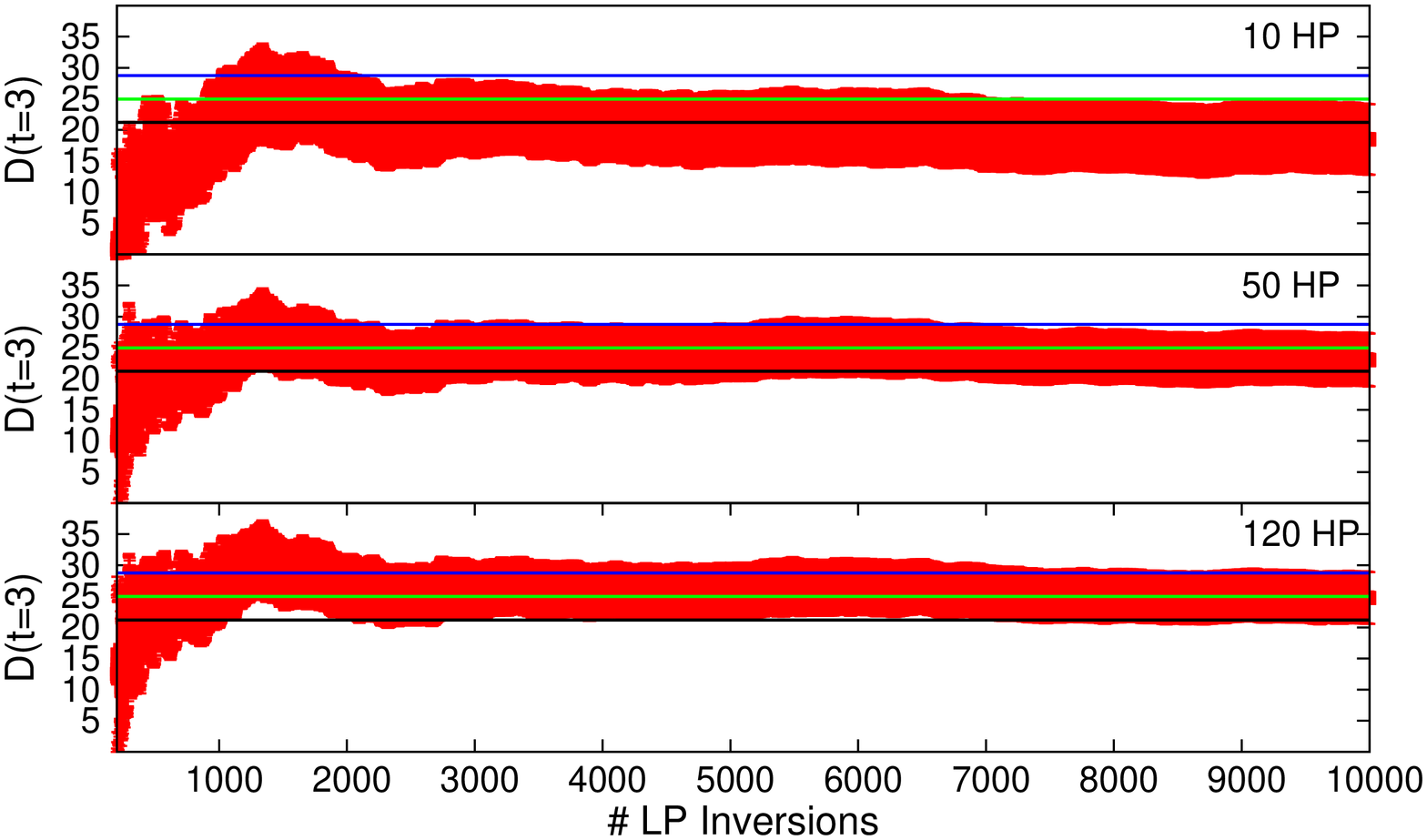}}
\end{minipage}
\caption{In the left panel we show the error on the disconnected and on the right 
the disconnected part of the $\eta^\prime$ correlator $D(t)$ at $t/a=3$ using the truncated solver method as a function of the number of low precision noise vectors  for: 10 (top), 50 (middle) and 120 (bottom) high precision noise vectors.  The lines show the mean value and error band of the exact result for
$D(t)$ at the same time slide.}
\label{fig:truncated}
\end{figure}

\begin{figure}[h!]
\begin{center}
{\includegraphics[angle=-90,width=0.7\linewidth]{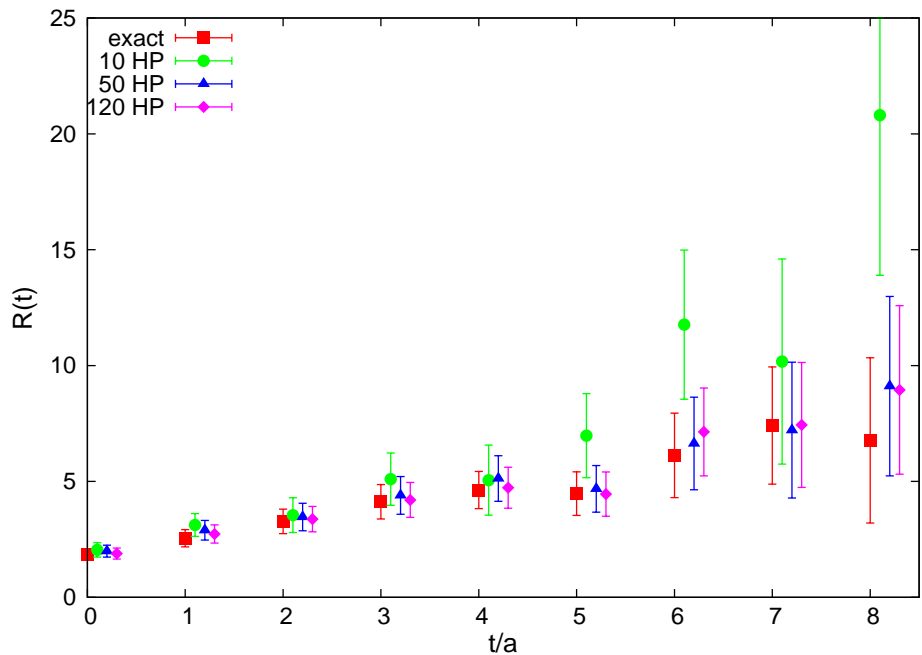}}
\end{center}
\caption{The ratio $R(t)=\frac{D(t)}{C_{\pi}(t)}$ computed using the truncated 
solver method and the exact approach. With the filled (red) squares we show the exact calculation and with the filled (green) circles, the filled (blue) triangles and the filled (magenta) rhombus  when using 10, 50 and 120  high precision noise   vectors, respectively.}
\label{fig:ratio}
\end{figure}
\begin{figure}[h!]
\begin{center}
{\includegraphics[angle=-90, width=0.7\linewidth]{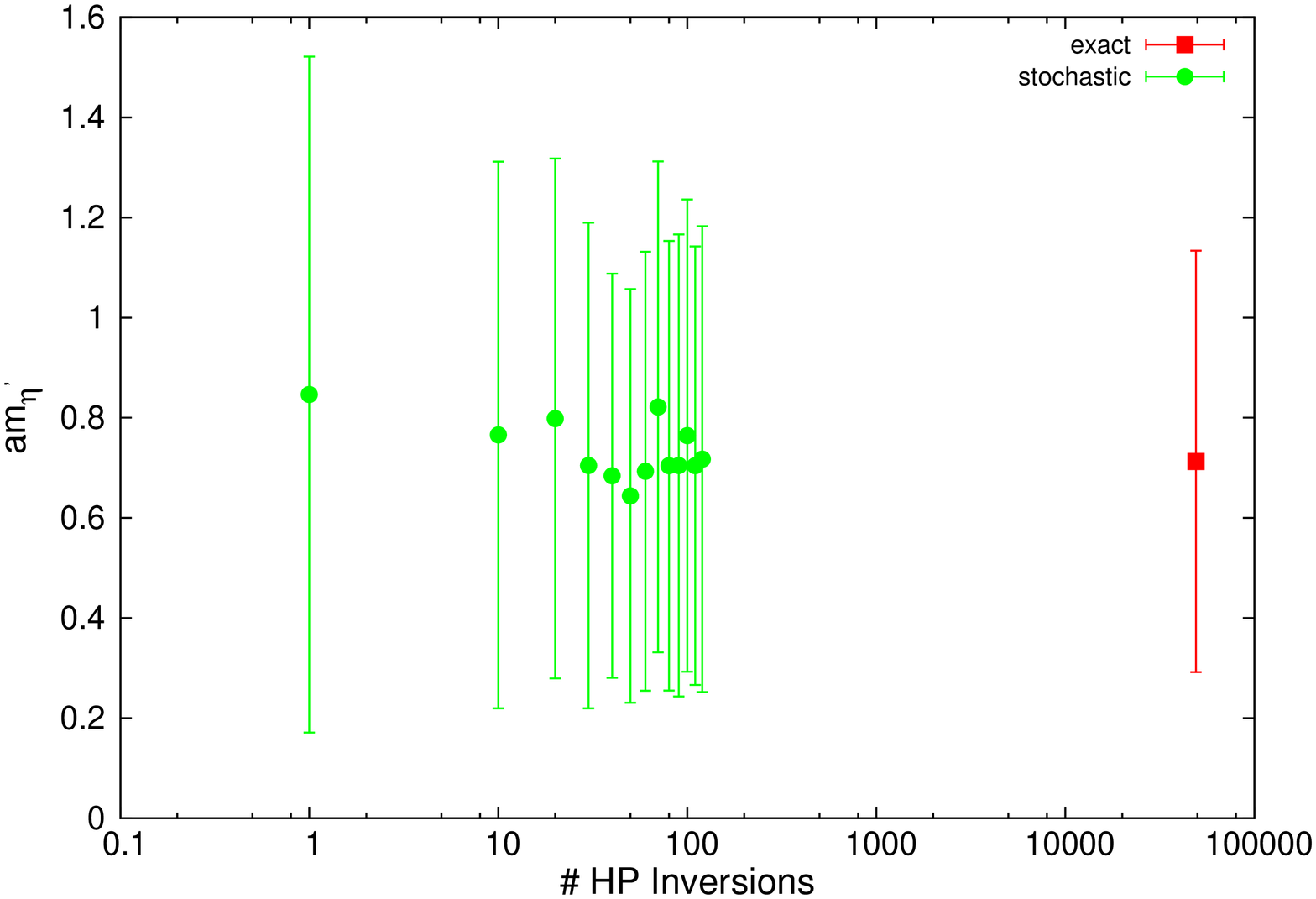}}
\end{center}
\caption{The mass of $\eta^\prime$ as  a function of computational cost (high precision inversions). The exact
result is shown with the filled (red) squares and the results of the stochastic truncated
solver method with the filled (green) circles as a function of the number of high precision vectors.}
\label{fig:eta_mass}
\end{figure}

For mesons on lattices with periodic boundary conditions the
two-point correlation function can be written as $C(t) \sim  e^{-mt}+e^{-m(T-t)} $ for $t$ large and $ t \ll T$, where $T$ is the lattice temporal extent.
 We can therefore analyze the ratio of the disconnected quark loop, $D(t)$, 
and connected correlation function, $C_{\pi}(t)$, to extract the flavor-singlet pseudo-scalar meson mass:
\begin{equation}
\frac{D(t)}{C_{\pi}(t)} \stackrel{t\rightarrow \infty}{\sim} A - B\frac{e^{-m_{\eta^\prime} t} + e^{-m_{\eta^\prime} (T-t)}}{e^{-m_\pi t} + e^{-m_\pi (T-t)}},
\label{eta mass}
\end{equation}
where $m_\pi$ and $m_{\eta^\prime}$ are the masses of the $\pi$ and $\eta^\prime$ mesons and $A$, $B$ are additional fit parameters.  
The pion mass $m_\pi$ can be determined separately by fitting the pion correlator to the $1\%$ level and used in Eq.~\ref{eta mass} leaving only 3 parameters in the fit function. 
Adopting this approach allows one to use 
 independent smearing of the connected and disconnected loops.

In the case of the exact evaluation, the only source of error
 comes from the statistical error of the gauge ensemble, and
therefore we will employ this fact to assess the results obtained using 
the different stochastic methods with color, spin, even-odd and cubic dilution.
In addition, we compare with  the truncated solver method where for the low precision inversions using BiCGstab we set the 
relative deviation i.e. $|Mx_i-b|/{|b|}$ to $10^{-2}$  and  for the high precision we set it to  $10^{-8}$, where $M$ is the Wilson Dirac operator, $x_i$  the solution vector after $i$ iterations and $b$ the source vector.

In Fig.~\ref{fig:compare dilution}
we show  results for the disconnected contributions $D(t)$ at $t/a=3$ as a function of the number of stochastic inversions  using color, spin, even-odd and cubic dilution. 
 In our
case the exact result is known and, since all results are computed
on the same gauge configurations,
 the stochastic evaluation should coincide with the result obtained from the exact evaluation of the all-to-all propagators in the 
limit of large enough noise vectors. Therefore, for the test case examined 
in this work, we show not only the error in the stochastic evaluation
but also the mean value, which
should coincide with  that of the exact 
evaluation. Using a different set of gauge configurations
will only reproduce the result within the statistical error
and therefore in practice one requires that the error obtained
in the stochastic evaluation converges to the  gauge error~\cite{Bali:2009hu}. 
As can be seen in Fig.~\ref{fig:compare dilution},
 the stochastic error remains almost unchanged after about 600-800 
 inversions for the four dilution schemes used.
The stochastic approach reproduces the mean value of the
 exact result as defined on a given set of gauge configurations 
and shown by the error band, in the limit of
a large  number of noise vectors.
 From this comparison we also conclude that
 the even-odd and cubic dilution schemes behave very similarly and therefore
in what follows we will show  results only for even-odd dilution, which is
most commonly used.  

In Fig.~\ref{fig:truncated} we show results for $D(t)$ at the same time slice, namely  $t/a=3$ as used  for the results in Fig.~\ref{fig:compare dilution}, but this time obtained with
  the truncated solver method. We display results as a function of the number
of low accuracy inversions increasing consecutively the number of high precision
noise vectors. Since the low precision is set to relative
precision $10^{-2}$ we only need about 10
BiCGStab iterations as compared to about 150 iterations for the high precision.
 As the number of HP vectors increases the stochastic error converges
to the statistical one  faster. However, for 10 HP vectors, although the
stochastic noise converges, the mean value remains
smaller than the exact result.
Therefore, one has to increase the HP vectors until also the mean value stabilizes. As can be seen, for 50 HP vectors  the mean value and the error
have converged
when the number of low precision vectors is above about
2500 i.e. when the number of high precision noise vectors is a few percentage the number of low precision vectors. This corresponds to a cost of about  215 HP inversions 
as compared to 
about 600 for the other stochastic methods. Therefore,
the truncated solver method is by far the most efficient in reproducing the exact result for the case of the loops with a $\gamma_5$ insertion.  

 Having established the preferred stochastic method for the evaluation
of the disconnected part we turn to the determination
of the $\eta^\prime$ mass in this $N_f=2$ theory.
 Given that $m_\pi$ is determined to the $1\%$ level it is clear that the gauge noise,  due to the disconnected loops, is large.

In Fig.~\ref{fig:ratio} we compare the exact  result for the ratio of disconnected to connected
to the results obtained using the truncated solver method.
As already mentioned, the ratio method allows us to use different smearing for
the disconnected and connected. We have therefore considered smeared
interpolating fields for the connected parts.
 As can be seen, 
the stochastic results  converge when 50 high precision noise vectors are used.  
Fitting the ratio $R(t)$  from $t=2a$ until  $t=8a$
we extract the mass of the $\eta^\prime$ shown in Fig.~\ref{fig:eta_mass}. 
 As can be seen, increasing the number of high precision noise vectors above 50
leaves the error and mean value unchanged. Therefore the truncated solver
method reproduces with low cost the 
the exact result. This enables computation of the loop at all time  slices
unlike in the case of the exact result where we  limit ourselves to 8 time
 slices. 
 The value we extract for the mass of the $\eta^\prime$ 
 is $am_{\eta^\prime}=0.54(10)$  or $m_{\eta^\prime}=1.17(22)$~GeV 
slightly higher than  the  value estimated by the
 SESAM collaboration  using these
configurations~\cite{Struckmann:2000bt}. 
Recently, the mass of the $\eta_2$ meson was studied using $N_f=2$ twisted mass fermions~\cite{Jansen:2008wv}. Within a pion mass range of about 500 MeV to about 300 MeV the dependence of the $\eta_2$ mass on the light quark mass was shown to be mild and
extrapolating to the physical point a value of 
0.865(65)(65)~GeV was obtained. Our value is thus in reasonable agreement 
taking into account
the higher light quark mass used in this study.

\section{Nucleon Electromagnetic form factors}
In order to
extract the nucleon electromagnetic form factors  we need to evaluate the nucleon matrix  
element $\langle N(p^\prime,s^\prime) | {j_\mu} | N(p,s) \rangle$, where  
$|N(p^\prime,s^\prime)\rangle$, $|N(p,s)\rangle$ are nucleon states with  
final momentum $p^\prime$ and spin $s^\prime$, and initial momentum  
$p$ and spin $s$.  
 The nucleon electromagnetic matrix element   
for real or  
virtual photons  can be decomposed in terms of the Dirac and Pauli form factors  $F_1$ and $F_2$ respectively:
\be  
 \langle \; N (p',s') \; | j^\mu | \; N (p,s) \rangle = 
  \bar{u} (p',s') \biggl[ \gamma^\mu F_1(q^2)   
+  \frac{i\sigma^{\mu\nu}q_\nu}{2m_N} F_2(q^2) \biggr] u(p,s) \; ,  
\label{Dirac ff}  
\ee  
where $q^2=(p^\prime-p)^2$,  
$ m_N$ is the nucleon mass and $E_N({\bf p})$ its energy.  
 $F_1(0)=1$ for the proton and zero for the neutron and
$F_2(0)$ measures the anomalous magnetic moment. $F_1$ and $F_2$ are connected to the  
electric, $G_E$, and magnetic, $G_M$, Sachs form factors by the relations  
\beq  
G_E(q^2)&=& F_1(q^2) + \frac{q^2}{(2m_N)^2} F_2(q^2)\nonumber \\  
G_M(q^2)&=& F_1(q^2) + F_2(q^2) \quad .  
\label{Sachs ff}  
\eeq  

An interpolating field for the proton is given by  
\be  
J(x) = \epsilon^{abc} \left[u^{a\, T}(x) \C\gamma_5 d^b(x)\right] u^c(x)  \quad.  
\ee  
 As described in section III, in order to increase the overlap with the nucleon state and  
decrease overlap with excited states we use Gaussian smeared quarks with APE 
smeared links.

In order to extract the  nucleon matrix element of Eq.~(\ref{Dirac ff})   
we need to calculate   
the two-point and three-point functions in Euclidean time defined by  
\hspace{-0.55cm}  
\beq  
\hspace{-0.55cm}G({\bf p}, t_f)\hspace{-0.15cm}&=&\hspace{-0.25cm}\sum_{{\bf x}_f} \, e^{-i{\bf x}_f \cdot {\bf p}}\,   
     {\Gamma_0^{\beta\alpha}}\, \langle {J_{\alpha}({\bf x}_f, t_f)}{\overline{J}_{\beta}({\bf 0},0)} \rangle \\  
\hspace{-0.5cm}  
G^\mu(\Gamma_\nu,{\bf q}, t) \hspace{-0.15cm}&=&\hspace{-0.25cm}\sum_{{\bf x}, {\bf x}_f} \, e^{i{\bf x}  
  \cdot {\bf q}}\,  \Gamma_\nu^{\beta\alpha}\, \langle  
{J_{\alpha}({\bf x}_f,t_f)} j^\mu({\bf x},t) {\overline{J}_{\beta}({\bf 0},0)}\rangle,  
\label{3-point}
\eeq  
where ${\Gamma_0}$ and ${\Gamma_k}$ are the projection matrices:  
\be  
{\Gamma_0} = \frac{1}{4}(\eins + \gamma_4)\,,\quad {\Gamma_k} =  
i{\Gamma_0} \gamma_5 \gamma_k\,.  
\ee  

\begin{figure}  
\begin{minipage}{0.42\linewidth}\vspace*{1.6cm}
 \includegraphics[width=\linewidth]{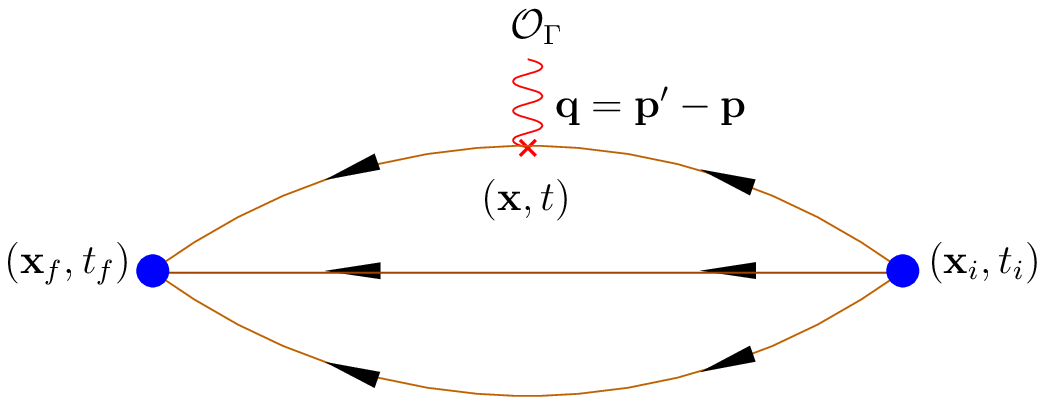}  
\end{minipage}\hfill
\begin{minipage}{0.42\linewidth}
 \includegraphics[width=\linewidth]{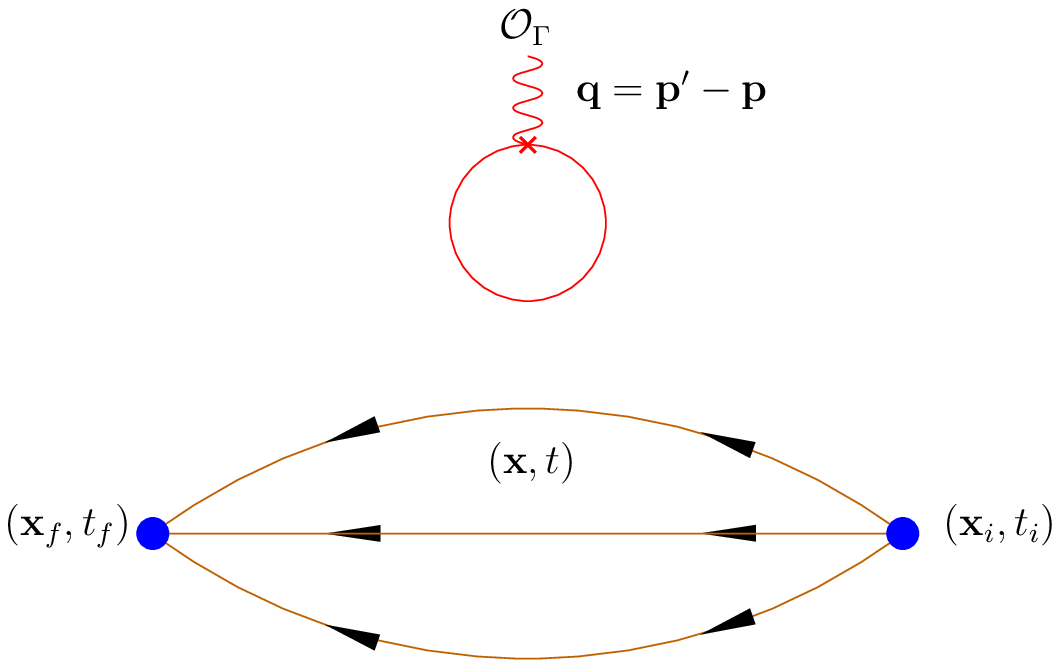}  
\end{minipage}
\caption{Left: Connected nucleon three-point function. Right: Disconnected nucleon three-point function.}  
\label{fig:FF diagrams}  
\end{figure}  

The kinematical setup that we used is illustrated in  
Fig.~\ref{fig:FF diagrams}: The creation (source) operator at  
time $t_i{=}0$ has fixed spatial position ${\bf x}_i {=} {\bf 0}$. The annihilation  
(sink) operator at a later time  
$t_f$ carries momentum ${\bf p}^\prime {=} 0$. The current couples to  
a quark at an intermediate time $t$ and carries the momentum ${\bf q}$.  
Translation invariance enforces ${\bf q}=-{\bf p}$  
for our kinematics.  
The form factors are calculated as a function of  
 $Q^2=-q^2>0$, which is the Euclidean momentum transfer squared.  
Provided the Euclidean times, $t$ and $t_f-t$ are large  
enough to filter the nucleon ground state,  
the time dependence  
of the Euclidean time evolution  
and the overlap factors cancel in the ratio  
\be  
R^{\mu}(\Gamma,{\bf q},t)= \frac{G^\mu(\Gamma,{\bf q},t) }{G({\bf 0},  
  t_f)}\ \sqrt{\frac{G({\bf p}, t_f-t)G({\bf 0},  t)G({\bf 0},  
    t_f)}{G({\bf 0}  , t_f-t)G({\bf p},t)G({\bf p},t_f)}}\,,  
\label{ratio}  
\ee  
 yielding a time-independent value  
\be  
\lim_{t_f-t\rightarrow \infty}\lim_{t\rightarrow \infty}R^{\mu}(\Gamma,{\bf q},t)=\Pi^\mu (\Gamma,{\bf q}) \,.  
\label{plateau}  
\ee  
We refer to the range of $t$-values where this asymptotic behavior is observed  
within our statistical precision as the plateau range.  
For this study, we use the local electromagnetic current
 $j^\mu (x)=\bar{\psi}(x)\gamma^\mu\psi(x)$ and take the renormalization constant $Z_V$  from Ref.~\cite{Becirevic:2005ta}.
We can extract the two Sachs form factors from the ratio of Eq.~(\ref{ratio}) by choosing  
appropriate combinations of the direction $\mu$ of the electromagnetic current  
  and projection matrices $\Gamma$.  
  
 Inclusion of  a complete set of hadronic
 states in the two- and three-point functions  
leads to the following expressions, written in Euclidean time:  
\be \Pi^{\mu=i} (\Gamma_k, {\bf q})  = C \frac{1}{2  
m_N} \epsilon_{ijk} \; q_j \; G_M (Q^2)   
\label{GM}  
\ee  
\be \Pi^{\mu=i} (\Gamma_0, {\bf q})  = C  
\frac{q_i}{2 m_N} \; G_E (Q^2)   
\label{GE123}  
\ee  
\be \Pi^{\mu=0} (\Gamma_0, {\bf q})  = C  
\frac{E_N +m_N}{2 m_N} \; G_E (Q^2)  \; ,  
\label{GE4}  
\ee  
where $C=  
\sqrt{\frac{2 m_N^2}{E_N(E_N + m_N)}}$ is a kinematical factor connected to the
normalization of the lattice states and the two-point functions entering in the 
ratio of Eq.~(\ref{ratio})~\cite{Alexandrou:2006ru}.

As schematically shown in Fig.~\ref{fig:FF diagrams},  the nucleon three-point function
 can be written in terms of a connected and a disconnected diagram.
 The connected diagram can 
be evaluated via the standard approach of computing the sequential propagator
through the sink. The
polarized matrix element given in Eq.~(\ref{GM}), from which  
 the magnetic form factor is determined,  requires an inversion for each
$\gamma_i$ in order that  we can calculate the matrix element for
 all momenta ${\bf q}$ in a symmetric way. For the small lattice
and heavy pion mass that we have in this study,  this can be done very fast. 
The goal of this work is the computation of the disconnected part,  given by
\beq
&\,&\langle J_\alpha(x_f) \vert j^\mu(x) \vert \bar{J}_\beta(x_i)\rangle\;_{\rm Disc.} \;=\; \frac{1}{3} Tr \left[ \gamma^\mu G(x,x) \right] \times \nonumber \\
&{}& \hspace*{-2.5cm}\epsilon^{abc} \epsilon^{a^\prime b^\prime c^\prime} (C \gamma_5 )_{\lambda\nu} (C \gamma_5)_{\lambda^\prime \nu^\prime} G_{\nu\nu^\prime}^{b b^\prime}(x_f,x_i) \left(  G_{\alpha\lambda^\prime}^{ca^\prime}(x_f,x_i) G_{\lambda\beta}^{ac^\prime}(x_f,x_i) - G_{\lambda\lambda^\prime}^{aa^\prime}(x_f,x_i) G_{\alpha\beta}^{c c^\prime}(x_f,x_i) \right).
\eeq 
As shown diagrammatically, this disconnected contribution  consists of the  
fermion loop multiplied by the nucleon two-point function. 
Using Eq.~(\ref{3-point})  one sees that one needs to perform the sum over the spatial coordinates of the current in the fermion loop  in order to obtain the nucleon matrix element, requiring knowledge of the all-to-all propagator.
Therefore the complexity lies in the evaluation
of the disconnected loop.

 \begin{figure}[h!]
\begin{minipage}{0.49\linewidth}
{\includegraphics[angle=-90, width=\linewidth]{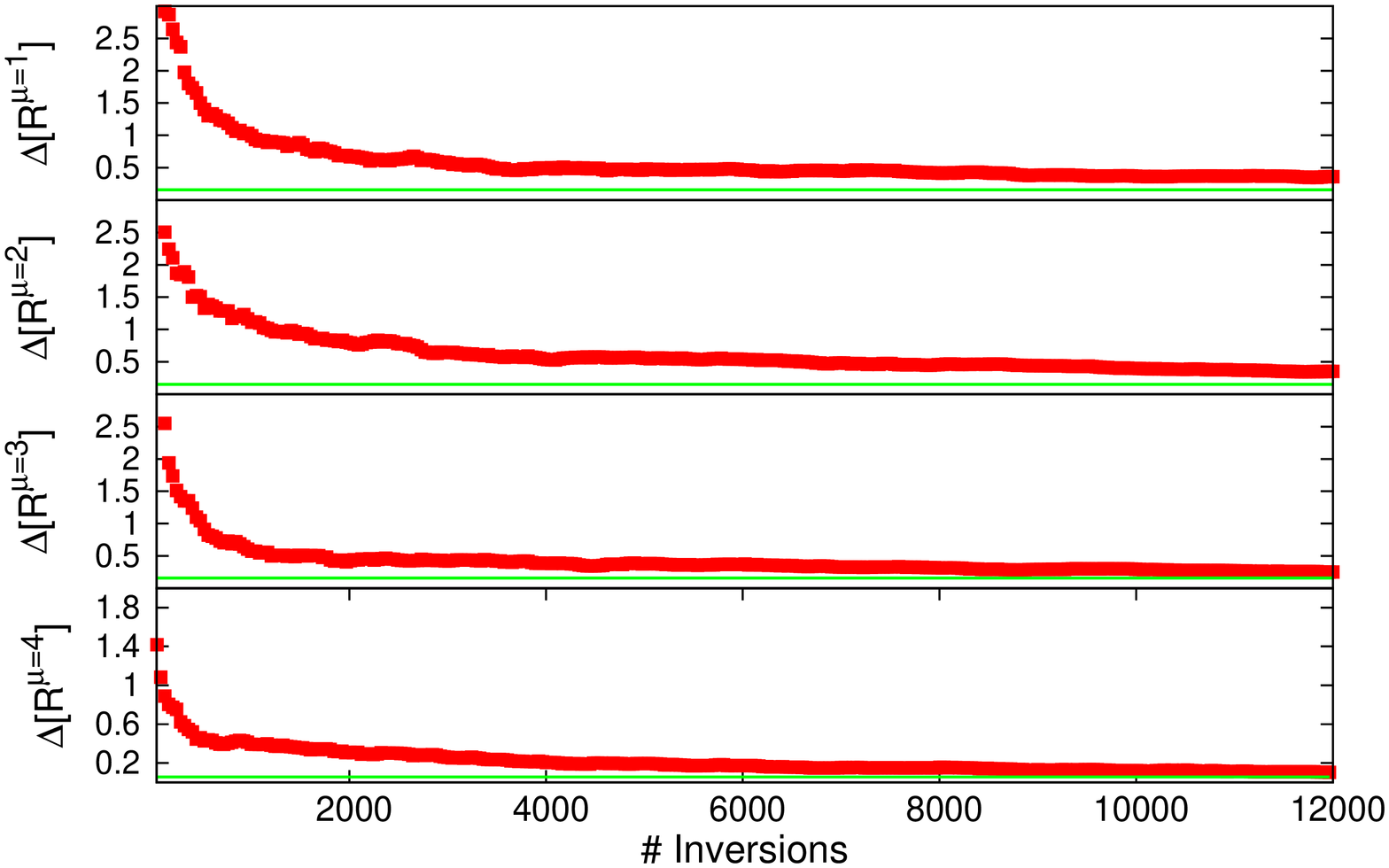}}
\end{minipage}
\begin{minipage}{0.49\linewidth}
{\includegraphics[angle=-90, width=\linewidth]{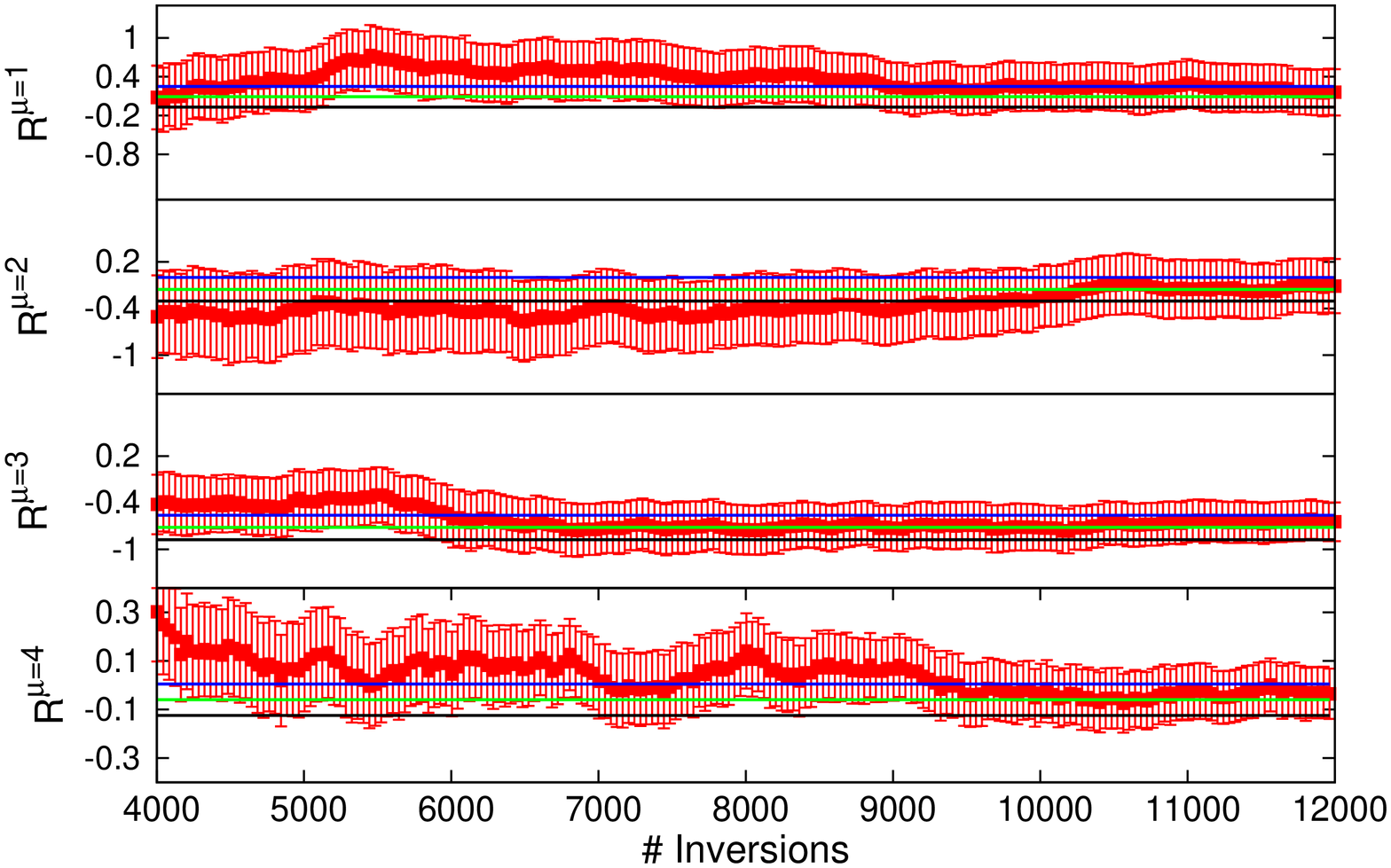}}
\end{minipage}
\caption{On the left we show the error  and on the right 
 the ratio given in Eq.~(\ref{ratio}) for the disconnected 
diagram using spin dilution. The lines indicate the exact value with its error
band. We show from top to bottom results: 
for $\gamma_1$ and $\vec{p}=(1,0,0)$; $\gamma_2$ and $\vec{p}=(0,1,0)$;  $\gamma_3$ and $\vec{p}=(0,0,1)$;  $\gamma_4$ and $\vec{p}=(1,0,0)$. In all cases the projection matrix $\Gamma_0$ is used and $t_f-t=4a$.} \label{fig:spin gammas}
\end{figure}

 \begin{figure}[h!]
\begin{minipage}{0.49\linewidth}
{\includegraphics[angle=-90, width=\linewidth]{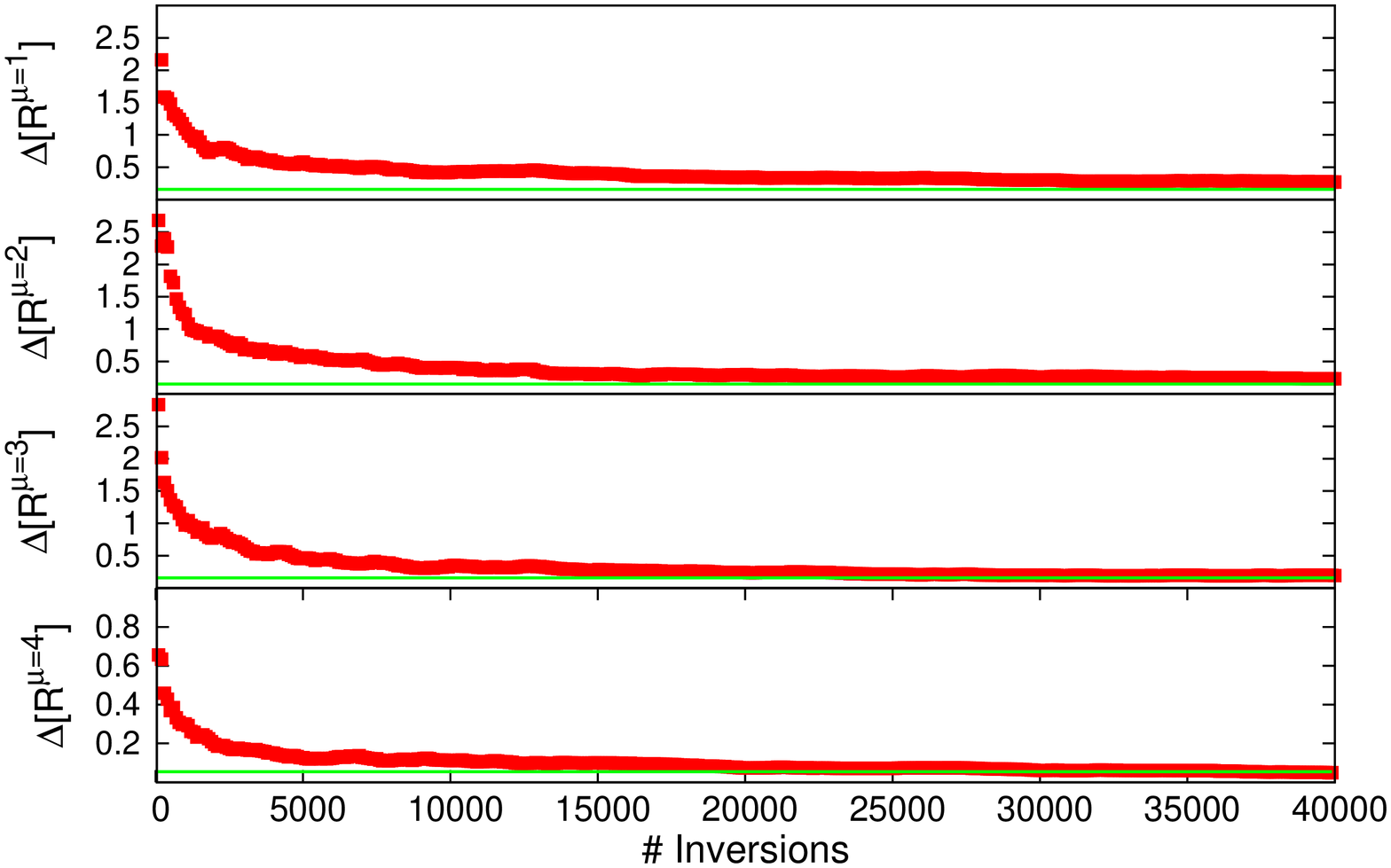}}
\end{minipage}
\begin{minipage}{0.49\linewidth}
{\includegraphics[angle=-90, width=\linewidth]{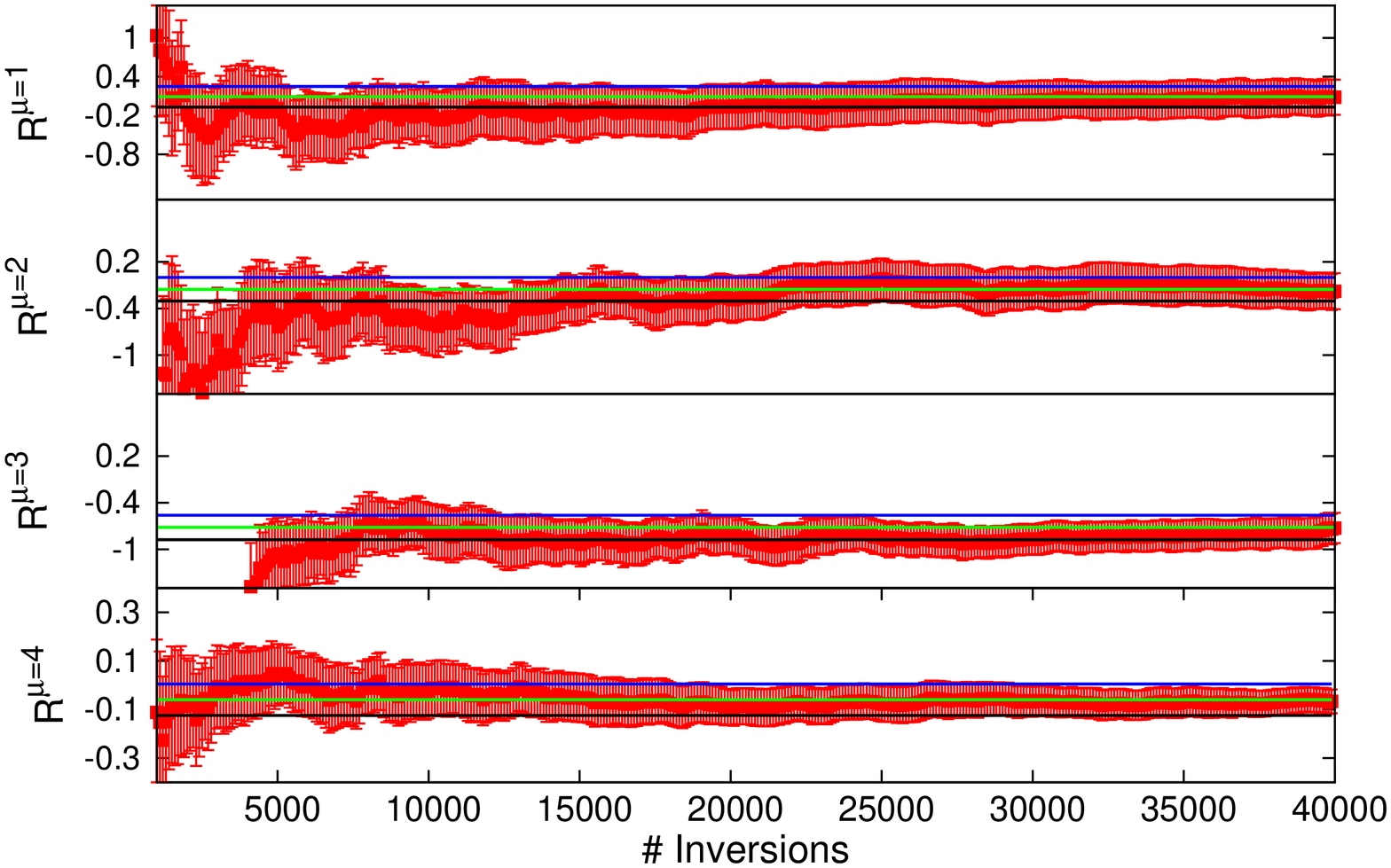}}
\end{minipage}
\caption{On the left we show the error and on the right
 the ratio given in Eq.~(\ref{ratio}) for the disconnected 
diagram using the truncated solver method versus the number of low precision noise vectors. 
The number of high precision noise vectors used are 1240 for which 
convergence is reached. The notation is the same as that in Fig.~\ref{fig:spin gammas}.} \label{fig:truncated gammas}
\end{figure}

In Fig.~\ref{fig:spin gammas} we compare exact results for the disconnected
diagram contributing to the ratio defined in Eq.~(\ref{ratio}) 
for various operator insertions with results obtained using spin dilution.
The behavior of the other stochastic methods, i.e. using color and even-odd dilution, is similar to that of spin dilution and 
therefore they are not shown. As can
be seen,  the number of stochastic vectors needed for convergence is large.
Even with about 25\% the cost of the exact evaluation the results have not fully converged. 
Therefore these stochastic dilution schemes are not very effective for
 calculating the loops with a $\gamma_\mu$ insertion.
 In Fig.~\ref{fig:truncated gammas} we make a similar comparison but
 using the truncated solver method. We show the results as a function of
the number of low precision noise vectors used. 
These results were obtained using of the order of $10^3$ high precision noise vectors, or about 2\% of the largest number of low precision vectors used.
As can be seen, convergence is achieved at much lower cost since,
although the number of low precision vectors used 
is of the same order as the number of inversions needed for the exact
evaluation, the cost in the former case is much lower.
 Therefore the truncated solver method is
by far the best choice  for the fermion loops entering the evaluation of the electromagnetic form factors.

\begin{figure}[h!]
  \begin{center}
    {\includegraphics[angle=-90, width=0.7\linewidth]{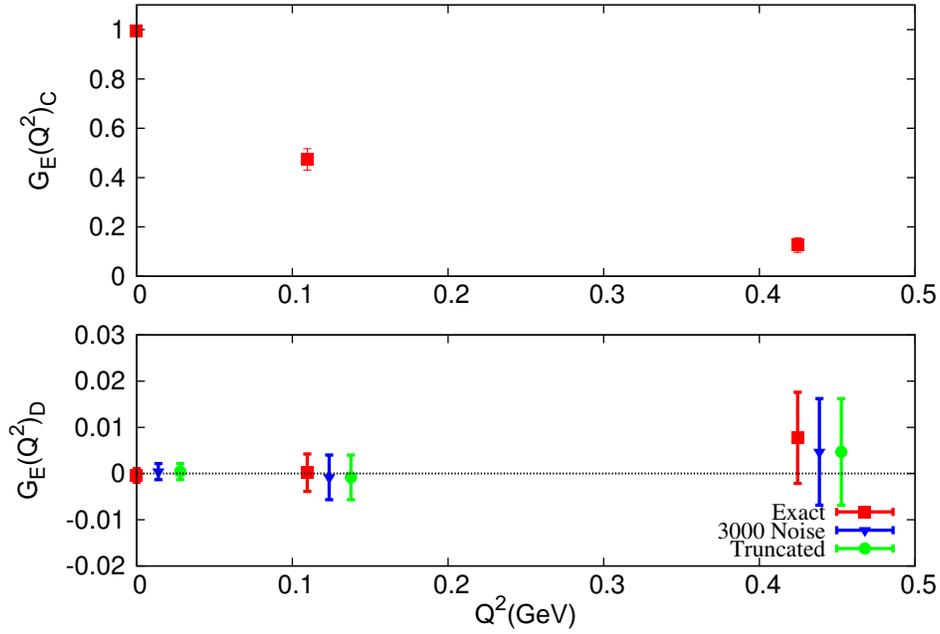}}
  \end{center}
  \caption{Results on the connected (top) and disconnected (bottom) part for the electric form factor. The exact result is shown with the filled (red) squares. The results are compared to spin dilution for 3000 noise vectors (or 12000 inversions) shown with the filled (blue) triangles and with the truncated solver method shown with the filled (green) circles.}
  \label{fig:electric ff}
\end{figure}

 \begin{figure}[h!]
\begin{center}
{\includegraphics[angle=-90, width=0.7\linewidth]{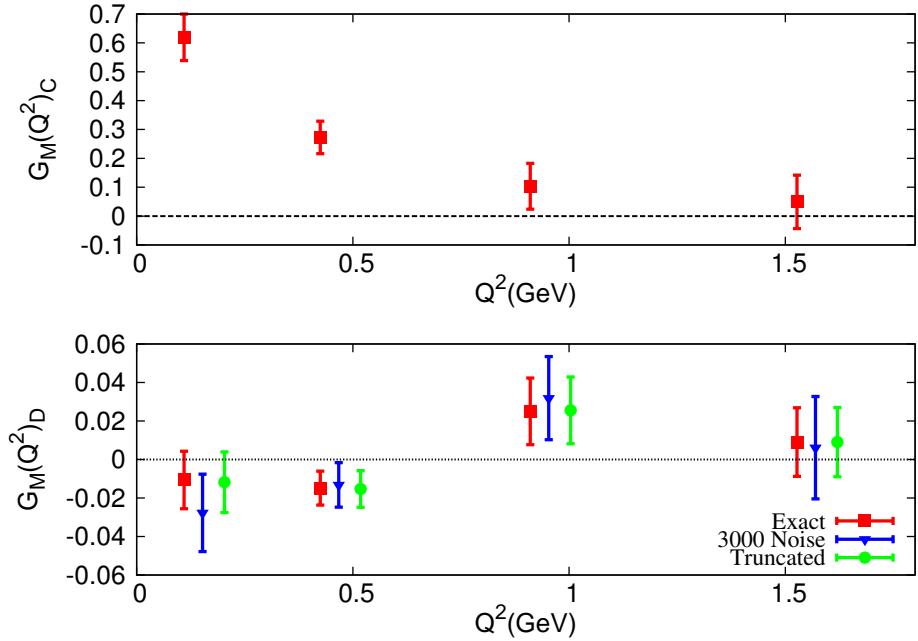}}
\end{center}
\caption{Results on the connected (top) and disconnected (bottom) part of the magnetic form factor. The notation is the same as that in Fig.~\ref{fig:electric ff}.}
\label{fig:magnetic ff}
\end{figure}

In Figs.~\ref{fig:electric ff} and \ref{fig:magnetic ff} we show the results
for the electric and magnetic form factors corresponding to the connected and disconnected contributions. 
As already noted, the results obtained using spin dilution have
not fully converged whereas the results obtained using the truncated solver
method are fully consistent with the exact evaluation.
 The fermion loops entering in the determination of the
electromagnetic  form factors are very noisy  and their contribution  at this pion mass is small.  
A similar conclusion is reached for the nucleon 
strange form factors in the study of Ref.~\cite{Babich:2010at} where the
fermion loops have similar quark mass as ours. 

\section{Nucleon scalar form factors}
Another important quantity where fermion loops may contribute significantly
is the nucleon $\sigma$-term given by
\be \sigma_l=m_l\langle N|\bar{u}u+\bar{d}d|N \rangle, \hspace*{0.5cm}m_l=\frac{1}{2}\left(m_u+m_d\right).
\label{sigma term}
\ee
 This is  proportional to the nucleon matrix element of the scalar quark density  at $Q^2 =0$. An equivalent quantity can be defined for the strange quark density. The precise knowledge of these quantities
are crucial as their value affects the magnitude 
of the dark matter cross sections on nuclear targets. Currently 
the uncertainty on their values represents the largest single uncertainty affecting the cross sections relevant in various super-symmetric models. It is therefore of the utmost importance to minimize the error on the $\sigma$- terms. In this
work we focus on the light quark sigma term $\sigma_l$, which 
is extracted from a chiral analysis of low energy pion-nucleus scattering data. However,  phenomenological analyses give somewhat different results~\cite{Giedt:2009mr}. For example
 an earlier analysis gave $\sigma_l = 45(8)$ MeV~\cite{Koch:1982pu}, whereas a more recent one gave $\sigma_l = 64(7)$ MeV~\cite{Pavan:2001wz}.

 \begin{figure}[h!]
\begin{minipage}{0.49\linewidth}
{\includegraphics[angle=-90, width=\linewidth]{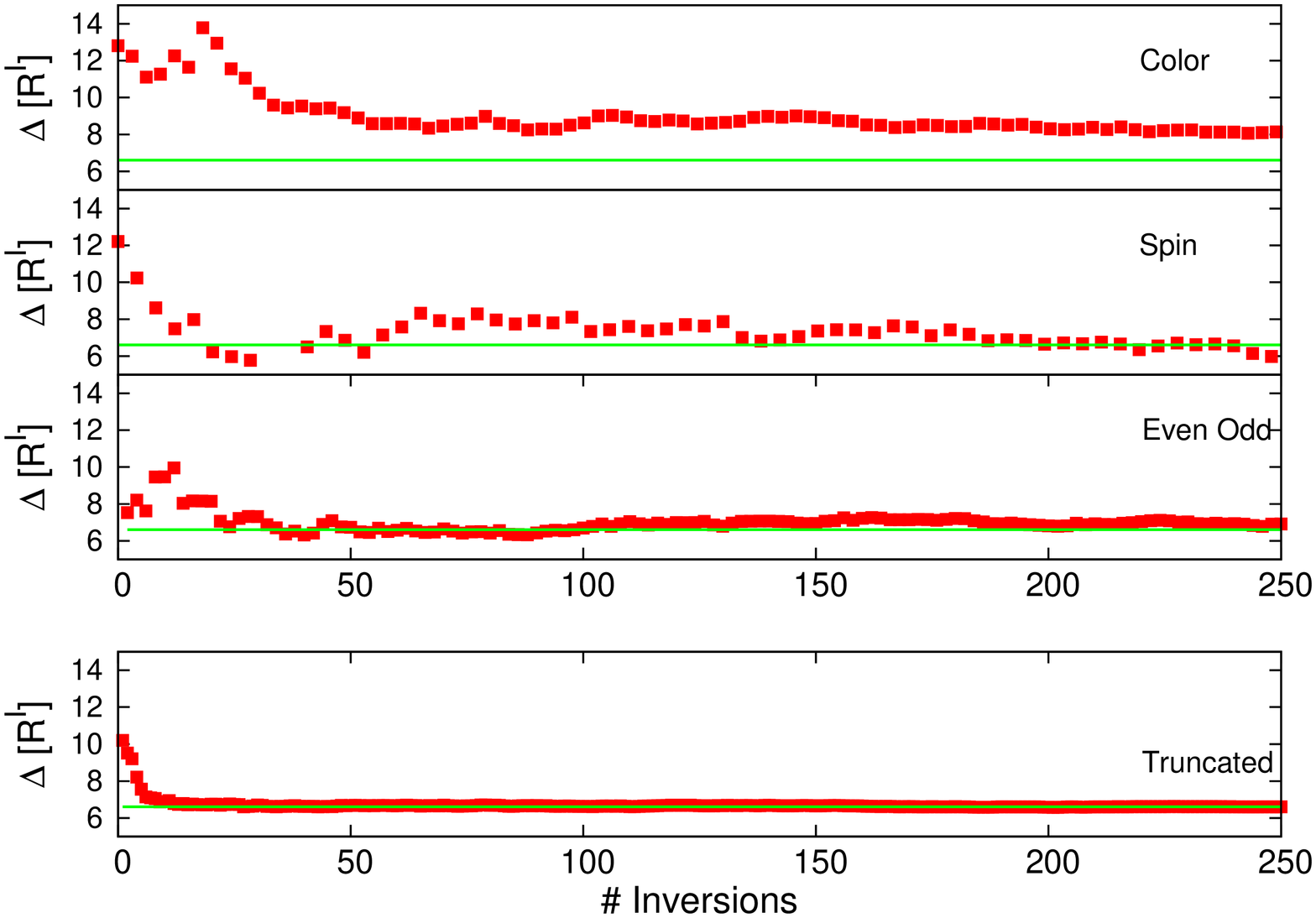}}
\end{minipage}
\begin{minipage}{0.49\linewidth}
{\includegraphics[angle=-90, width=\linewidth]{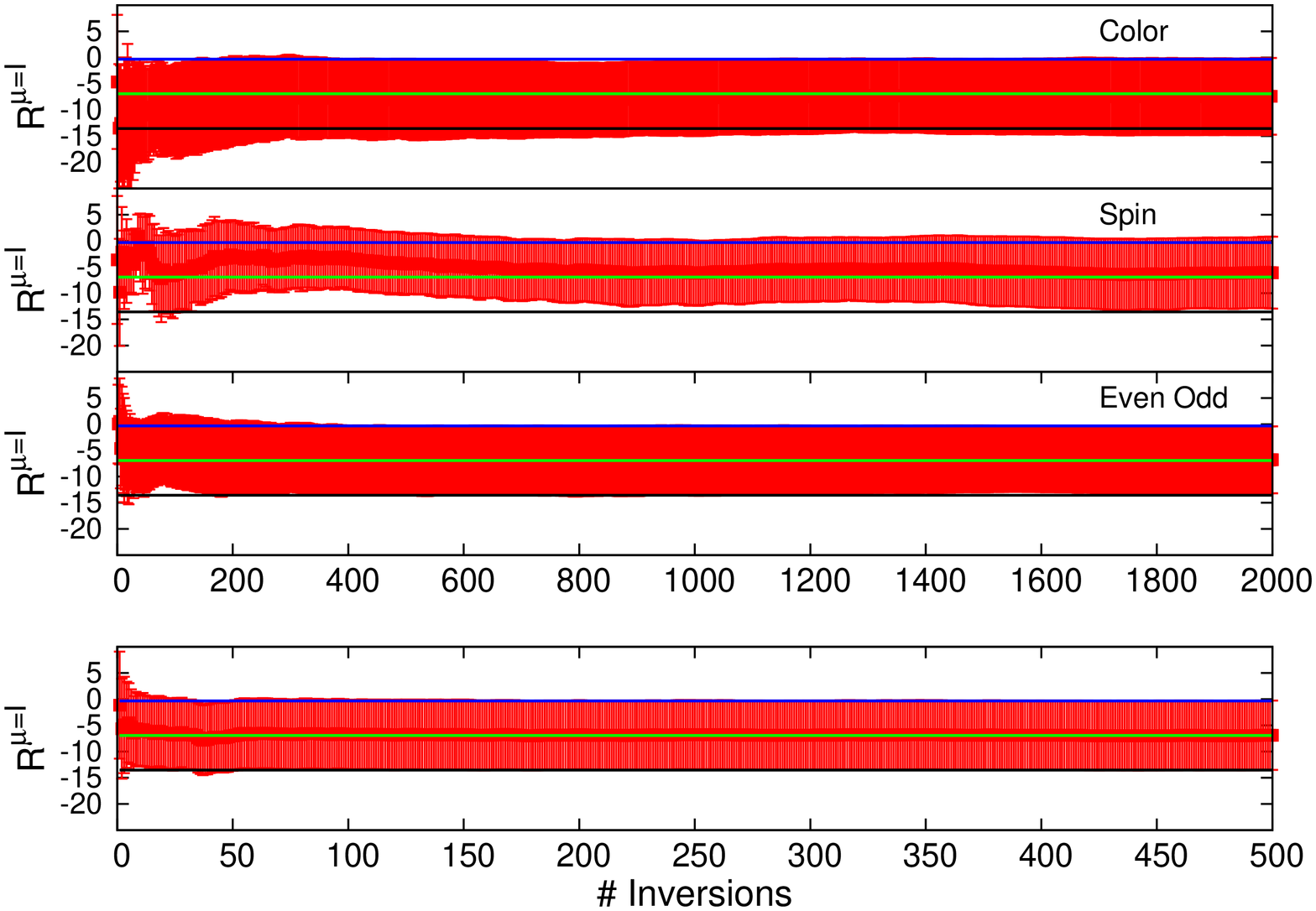}}
\end{minipage}
\caption{Comparison of the three dilution methods (color, spin, odd-even)  (top) and the truncated solver method (bottom) for the scalar current. On the left panel we show
the error and on the right panel the ratio given in Eq.~\ref{ratio}.
  In the case of the truncated solver method, we plot against the number of low-precision inversions, 
  while the high precision inversions are fixed to about 2\% of the maximum
 number of low-precision vectors used.}
\label{fig:compare_unity}
\end{figure}

 \begin{figure}[h!]
\begin{center}
{\includegraphics[angle=-90, width=0.7\linewidth]{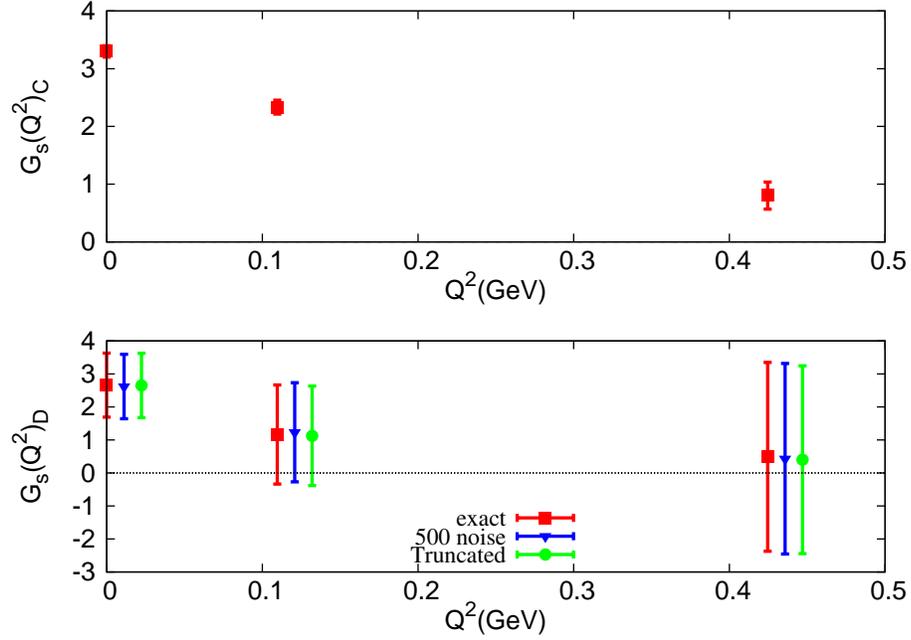}}
\end{center}
\caption{The connected (top) and the disconnected (bottom) contributions of the scalar form factor.
The exact result is shown with the filled (red) squares. The results are compared to spin dilution for 500 noise vectors (or 2000 inversions) shown with the filled (blue) triangles and with the truncated solver method shown with the filled (green) circles.}
\label{fig:scalar_form_factor}
\end{figure}
 \begin{figure}[h!]
\begin{center}
{\includegraphics[width=0.5\linewidth]{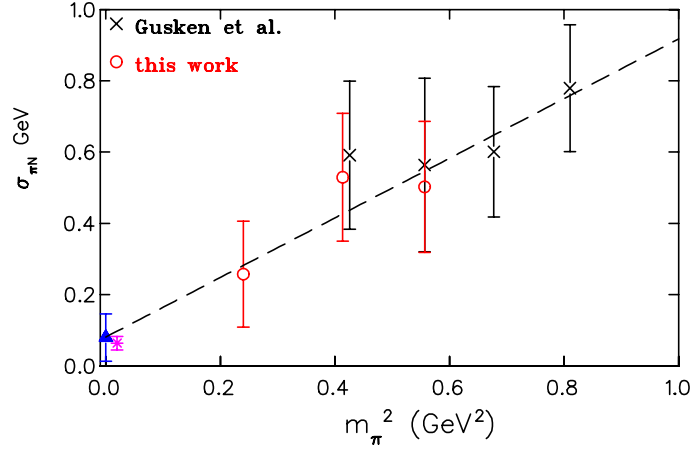}}
\end{center}
\caption{The nucleon $\sigma$-term as a function of the pion mass squared.
The crosses show results derived from the values of the
three-point function calculated in  Ref.~\cite{Gusken:1998wy}, whereas
the open circles are the results of this work. The line is a linear extrapolation to the chiral limit, giving
the value shown by the filled triangle. The asterisk shows the
 value of $\sigma_l$ extracted from a recent phenomenological
analysis~\cite{Pavan:2001wz} with a systematic error that shows  
the deviation of this value from an earlier phenomenological  determination given in Ref.~\cite{Koch:1982pu}.
}
\label{fig:sigma}
\end{figure}

An indirect method to extract the nucleon $\sigma$-terms from lattice QCD without
computing the disconnected diagram  is to evaluate the 
dependence of the nucleon mass on the light and strange quark mass~\cite{Alexandrou:2009qu,Alexandrou:2008tn,Young:2009zb,WalkerLoud:2008bp,Ohki:2008ff,Aoki:2008sm,Durr:2010ni}.
However, recently stochastic methods have been applied to evaluate 
directly the nucleon $\sigma$-terms~\cite{Babich:2010at,Bali:2009hu}. Therefore it is interesting to 
compare the various approaches.
In Fig.~\ref{fig:compare_unity} we perform a comparison of the commonly used dilution schemes  for the evaluation
of the scalar operator. 
As can be seen, the convergence for this operator is much better than in the case of the electromagnetic current, presented in the previous section.
Namely, for the scalar case, we converge to the exact value with as little as 500 noise vectors as compared to the case of the  
electromagnetic current where  even using as much as 12 000 noise vectors, which corresponds to
about 1/4 of the number of inversions needed for the exact evaluation, the stochastic error has not converged.
Therefore it is a matter of taste
which dilution scheme one employs in the evaluation of the  nucleon matrix element of the scalar quark density and $\sigma$-term. The truncated solver method convergences very fast,  at only a fraction of the computational cost of the other methods, 
 and proves also here to be the most efficient.  In Fig.~\ref{fig:scalar_form_factor} we show results on the nucleon scalar density as a function of the momentum transfer square, both for the connected and disconnected contributions. As can be seen,   the exact result for the disconnected contribution is reproduced using either noise vectors with spin dilution or the truncated solver method. 

In order to calculate $\sigma_l$ one needs to evaluate, besides the 
connected and disconnected contributions, the quark mass $m_l$.
Using the axial Ward-Takahashi identity 
\be 
 \partial^\mu A_\mu^a=2m_qP^a \quad,
\label{ward}
\ee
we can extract the  quark mass  by taking 
the matrix element of Eq.~(\ref{ward}) between a zero momentum
pion state and
the vacuum:
\be 
m_q = \frac{m_\pi<0|A_0^a|\pi^a(0)>}{2<0|P^a|\pi^a(0)>}
 \quad.
\label{quark mass}
\ee 
In order to obtained the renormalized quark mass one
needs the renormalization constant for the axial-vector current  $Z_A$ and
for the pseudoscalar current $Z_P$, whereas for the scalar density one
needs $Z_S$, which can be
taken from Ref.~\cite{Becirevic:2005ta}. 
However, the $\sigma$-term is renormalization group invariant and  therefore no renormalization is needed.

We find that the $\sigma$-term due to the connected diagram is
 $\sigma_l^C=0.278(8)$~GeV  and due to the disconnected 
$\sigma_l^D=0.224(82)$~GeV, giving $\sigma_l=0.50(8)$~GeV at pion mass $\sim 750$~MeV. 
This value is in agreement with an estimate made on the
same ensemble in Ref.~\cite{Gusken:1998wy} if one uses their
values of the connected and disconnected three-point function and the PCAC mass
instead of the naive quark mass used there. 
In order to investigate the pion mass dependence of the $\sigma$-term, we 
performed the calculation using the truncated solver method for the disconnected
contribution at two lighter quark masses, namely at $\kappa=0.1575$ and $\kappa=0.1580$ on a lattice of size $24^3\times 48$~\cite{Orth:2005kq}. The values of the quark mass
for this set of $\kappa$ values are taken  from Ref.~\cite{Orth:2005kq}.
The results are shown in Fig.~\ref{fig:sigma} where we also included
results from Ref.~\cite{Gusken:1998wy} but using the values of the  PCAC mass
given in  Ref.~\cite{Orth:2005kq}. Extrapolating linearly in the quark mass
one obtains a value at the physical point that is in  agreement with phenomenological
estimates of this quantity, albeit with a large statistical error.


\section{Conclusions}
The focus of this study is to investigate the stochastic techniques that are
commonly being applied to compute fermion loops by comparing to the exact
evaluation. Therefore, we
perform an exact evaluation using GPUs for a relatively small lattice of 
$16^3\times 32$ and $N_f=2$ Wilson fermions corresponding to a 
pion mass of about 750 MeV.   We  consider fermion
loops with the $\bar{\psi} \gamma_\mu \psi$ operator which are relevant
for the electromagnetic current, flavor singlet operators relevant for the
calculation of the $\eta^\prime$ mass and scalar operators relevant for
the calculation of the $\sigma$-term.

Comparing color, spin and two spatial  types of dilution schemes the conclusion is that they perform similarly. For the scalar operator, the convergence of these
dilution schemes is much faster and one 
needs an order of magnitude less noise vectors as compared
to the number needed when  there is a $\gamma$-insertion in the loop. Comparing the aforementioned 
dilution schemes with the truncated 
solver method the conclusion is that the latter is by far the most efficient. 
This conclusion holds for fermion loops involving light quarks.
A possible extension of this work will be  the
calculation of all-to-all propagators involved in other processes
 as for example
in the evaluation of three-point diagrams. Such a study was carried out
in the case of the semileptonic form factor of the D-meson~\cite{Evans:2010tg} and it would be
interesting to examine similar techniques  in the case of the nucleon.

\section{Acknowledgment}
C.A. acknowledges partial support by the Research Executive Agency (REA) of the European Union under Grant Agreement number PITN-GA-2009-238353 (ITN STRONGnet) and A. O'C. and A.S.  by the Cyprus Research Foundation under grant $\rm \Pi PO\Sigma E\Lambda K Y\Sigma H$/$\rm \Pi PONE$ 0308/09.
Computations partly utilized the Lincoln GPU cluster  provided  by the National Center for Supercomputing Applications.
We would like to thank A. W. Thomas for very constructive comments on the nucleon $\sigma$-terms.

\bibliography{ref}

\begin{thebibliography}{36}
\expandafter\ifx\csname natexlab\endcsname\relax\def\natexlab#1{#1}\fi
\expandafter\ifx\csname bibnamefont\endcsname\relax
  \def\bibnamefont#1{#1}\fi
\expandafter\ifx\csname bibfnamefont\endcsname\relax
  \def\bibfnamefont#1{#1}\fi
\expandafter\ifx\csname citenamefont\endcsname\relax
  \def\citenamefont#1{#1}\fi
\expandafter\ifx\csname url\endcsname\relax
  \def\url#1{\texttt{#1}}\fi
\expandafter\ifx\csname urlprefix\endcsname\relax\def\urlprefix{URL }\fi
\providecommand{\bibinfo}[2]{#2}
\providecommand{\eprint}[2][]{\url{#2}}

\bibitem[{\citenamefont{Aniol et~al.}(2006)}]{Aniol:2005zf}
\bibinfo{author}{\bibfnamefont{K.}~\bibnamefont{Aniol}} \bibnamefont{et~al.}
  (\bibinfo{collaboration}{HAPPEX Collaboration}),
  \bibinfo{journal}{Phys.Rev.Lett.} \textbf{\bibinfo{volume}{96}},
  \bibinfo{pages}{022003} (\bibinfo{year}{2006}), \eprint{nucl-ex/0506010}.

\bibitem[{\citenamefont{Armstrong et~al.}(2005)}]{Armstrong:2005hs}
\bibinfo{author}{\bibfnamefont{D.}~\bibnamefont{Armstrong}}
  \bibnamefont{et~al.} (\bibinfo{collaboration}{G0 Collaboration}),
  \bibinfo{journal}{Phys.Rev.Lett.} \textbf{\bibinfo{volume}{95}},
  \bibinfo{pages}{092001} (\bibinfo{year}{2005}), \eprint{nucl-ex/0506021}.

\bibitem[{\citenamefont{Maas et~al.}(2005)\citenamefont{Maas, Aulenbacher,
  Baunack, Capozza, Diefenbach et~al.}}]{Maas:2004dh}
\bibinfo{author}{\bibfnamefont{F.}~\bibnamefont{Maas}},
  \bibinfo{author}{\bibfnamefont{K.}~\bibnamefont{Aulenbacher}},
  \bibinfo{author}{\bibfnamefont{S.}~\bibnamefont{Baunack}},
  \bibinfo{author}{\bibfnamefont{L.}~\bibnamefont{Capozza}},
  \bibinfo{author}{\bibfnamefont{J.}~\bibnamefont{Diefenbach}},
  \bibnamefont{et~al.}, \bibinfo{journal}{Phys.Rev.Lett.}
  \textbf{\bibinfo{volume}{94}}, \bibinfo{pages}{152001}
  (\bibinfo{year}{2005}), \eprint{nucl-ex/0412030}.

\bibitem[{\citenamefont{Maas et~al.}(2004)}]{Maas:2004ta}
\bibinfo{author}{\bibfnamefont{F.}~\bibnamefont{Maas}} \bibnamefont{et~al.}
  (\bibinfo{collaboration}{A4 Collaboration}),
  \bibinfo{journal}{Phys.Rev.Lett.} \textbf{\bibinfo{volume}{93}},
  \bibinfo{pages}{022002} (\bibinfo{year}{2004}), \eprint{nucl-ex/0401019}.

\bibitem[{\citenamefont{Spayde et~al.}(2004)}]{Spayde:2003nr}
\bibinfo{author}{\bibfnamefont{D.}~\bibnamefont{Spayde}} \bibnamefont{et~al.}
  (\bibinfo{collaboration}{SAMPLE Collaboration}),
  \bibinfo{journal}{Phys.Lett.} \textbf{\bibinfo{volume}{B583}},
  \bibinfo{pages}{79} (\bibinfo{year}{2004}), \eprint{nucl-ex/0312016}.

\bibitem[{\citenamefont{Leinweber et~al.}(2005)\citenamefont{Leinweber,
  Boinepalli, Cloet, Thomas, Williams et~al.}}]{Leinweber:2004tc}
\bibinfo{author}{\bibfnamefont{D.~B.} \bibnamefont{Leinweber}},
  \bibinfo{author}{\bibfnamefont{S.}~\bibnamefont{Boinepalli}},
  \bibinfo{author}{\bibfnamefont{I.}~\bibnamefont{Cloet}},
  \bibinfo{author}{\bibfnamefont{A.~W.} \bibnamefont{Thomas}},
  \bibinfo{author}{\bibfnamefont{A.~G.} \bibnamefont{Williams}},
  \bibnamefont{et~al.}, \bibinfo{journal}{Phys.Rev.Lett.}
  \textbf{\bibinfo{volume}{94}}, \bibinfo{pages}{212001}
  (\bibinfo{year}{2005}), \eprint{hep-lat/0406002}.

\bibitem[{\citenamefont{Leinweber et~al.}(2006)\citenamefont{Leinweber,
  Boinepalli, Thomas, Wang, Williams et~al.}}]{Leinweber:2006ug}
\bibinfo{author}{\bibfnamefont{D.~B.} \bibnamefont{Leinweber}},
  \bibinfo{author}{\bibfnamefont{S.}~\bibnamefont{Boinepalli}},
  \bibinfo{author}{\bibfnamefont{A.~W.} \bibnamefont{Thomas}},
  \bibinfo{author}{\bibfnamefont{P.}~\bibnamefont{Wang}},
  \bibinfo{author}{\bibfnamefont{A.~G.} \bibnamefont{Williams}},
  \bibnamefont{et~al.}, \bibinfo{journal}{Phys.Rev.Lett.}
  \textbf{\bibinfo{volume}{97}}, \bibinfo{pages}{022001}
  (\bibinfo{year}{2006}), \eprint{hep-lat/0601025}.

\bibitem[{\citenamefont{Jansen et~al.}(2008)\citenamefont{Jansen, Michael, and
  Urbach}}]{Jansen:2008wv}
\bibinfo{author}{\bibfnamefont{K.}~\bibnamefont{Jansen}},
  \bibinfo{author}{\bibfnamefont{C.}~\bibnamefont{Michael}}, \bibnamefont{and}
  \bibinfo{author}{\bibfnamefont{C.}~\bibnamefont{Urbach}}
  (\bibinfo{collaboration}{ETM Collaboration}), \bibinfo{journal}{Eur.Phys.J.}
  \textbf{\bibinfo{volume}{C58}}, \bibinfo{pages}{261} (\bibinfo{year}{2008}),
  \eprint{0804.3871}.

\bibitem[{\citenamefont{Bali et~al.}(2010)\citenamefont{Bali, Collins, and
  Schafer}}]{Bali:2009hu}
\bibinfo{author}{\bibfnamefont{G.~S.} \bibnamefont{Bali}},
  \bibinfo{author}{\bibfnamefont{S.}~\bibnamefont{Collins}}, \bibnamefont{and}
  \bibinfo{author}{\bibfnamefont{A.}~\bibnamefont{Schafer}},
  \bibinfo{journal}{Comput.Phys.Commun.} \textbf{\bibinfo{volume}{181}},
  \bibinfo{pages}{1570} (\bibinfo{year}{2010}), \eprint{0910.3970}.

\bibitem[{\citenamefont{Doi et~al.}(2009)\citenamefont{Doi, Deka, Dong, Draper,
  Liu et~al.}}]{Doi:2009sq}
\bibinfo{author}{\bibfnamefont{T.}~\bibnamefont{Doi}},
  \bibinfo{author}{\bibfnamefont{M.}~\bibnamefont{Deka}},
  \bibinfo{author}{\bibfnamefont{S.-J.} \bibnamefont{Dong}},
  \bibinfo{author}{\bibfnamefont{T.}~\bibnamefont{Draper}},
  \bibinfo{author}{\bibfnamefont{K.-F.} \bibnamefont{Liu}},
  \bibnamefont{et~al.}, \bibinfo{journal}{Phys.Rev.}
  \textbf{\bibinfo{volume}{D80}}, \bibinfo{pages}{094503}
  (\bibinfo{year}{2009}), \eprint{0903.3232}.

\bibitem[{\citenamefont{Babich et~al.}(2010)\citenamefont{Babich, Brower,
  Clark, Fleming, Osborn et~al.}}]{Babich:2010at}
\bibinfo{author}{\bibfnamefont{R.}~\bibnamefont{Babich}},
  \bibinfo{author}{\bibfnamefont{R.~C.} \bibnamefont{Brower}},
  \bibinfo{author}{\bibfnamefont{M.~A.} \bibnamefont{Clark}},
  \bibinfo{author}{\bibfnamefont{G.~T.} \bibnamefont{Fleming}},
  \bibinfo{author}{\bibfnamefont{J.~C.} \bibnamefont{Osborn}},
  \bibnamefont{et~al.} (\bibinfo{year}{2010}), \eprint{1012.0562}.

\bibitem[{\citenamefont{Takeda et~al.}(2011)}]{Takeda:2010cw}
\bibinfo{author}{\bibfnamefont{K.}~\bibnamefont{Takeda}} \bibnamefont{et~al.}
  (\bibinfo{collaboration}{JLQCD collaboration}), \bibinfo{journal}{Phys.Rev.}
  \textbf{\bibinfo{volume}{D83}}, \bibinfo{pages}{114506}
  (\bibinfo{year}{2011}), \eprint{1011.1964}.

\bibitem[{\citenamefont{Feng et~al.}(2011)\citenamefont{Feng, Jansen,
  Petschlies, and Renner}}]{Feng:2011zk}
\bibinfo{author}{\bibfnamefont{X.}~\bibnamefont{Feng}},
  \bibinfo{author}{\bibfnamefont{K.}~\bibnamefont{Jansen}},
  \bibinfo{author}{\bibfnamefont{M.}~\bibnamefont{Petschlies}},
  \bibnamefont{and} \bibinfo{author}{\bibfnamefont{D.~B.} \bibnamefont{Renner}}
  (\bibinfo{year}{2011}), \eprint{1103.4818}.

\bibitem[{\citenamefont{Barros et~al.}(2008)\citenamefont{Barros, Babich,
  Brower, Clark, and Rebbi}}]{Barros:2008rd}
\bibinfo{author}{\bibfnamefont{K.}~\bibnamefont{Barros}},
  \bibinfo{author}{\bibfnamefont{R.}~\bibnamefont{Babich}},
  \bibinfo{author}{\bibfnamefont{R.}~\bibnamefont{Brower}},
  \bibinfo{author}{\bibfnamefont{M.~A.} \bibnamefont{Clark}}, \bibnamefont{and}
  \bibinfo{author}{\bibfnamefont{C.}~\bibnamefont{Rebbi}},
  \bibinfo{journal}{PoS} \textbf{\bibinfo{volume}{LATTICE2008}},
  \bibinfo{pages}{045} (\bibinfo{year}{2008}), \eprint{0810.5365}.

\bibitem[{\citenamefont{Alexandrou et~al.}(2010)\citenamefont{Alexandrou,
  Christaras, O'Cais, and Strelchenko}}]{Alexandrou:2010jr}
\bibinfo{author}{\bibfnamefont{C.}~\bibnamefont{Alexandrou}},
  \bibinfo{author}{\bibfnamefont{D.}~\bibnamefont{Christaras}},
  \bibinfo{author}{\bibfnamefont{A.}~\bibnamefont{O'Cais}}, \bibnamefont{and}
  \bibinfo{author}{\bibfnamefont{A.}~\bibnamefont{Strelchenko}},
  \bibinfo{journal}{PoS} \textbf{\bibinfo{volume}{LATTICE2010}},
  \bibinfo{pages}{035} (\bibinfo{year}{2010}), \eprint{1012.5168}.

\bibitem[{\citenamefont{Hoeber et~al.}(1998)}]{Hoeber:1997rg}
\bibinfo{author}{\bibfnamefont{H.}~\bibnamefont{Hoeber}} \bibnamefont{et~al.}
  (\bibinfo{collaboration}{TXL Collaboration, TkL Collaboration}),
  \bibinfo{journal}{Nucl.Phys.Proc.Suppl.} \textbf{\bibinfo{volume}{63}},
  \bibinfo{pages}{218} (\bibinfo{year}{1998}), \eprint{hep-lat/9709137}.

\bibitem[{\citenamefont{Orth et~al.}(2005)\citenamefont{Orth, Lippert, and
  Schilling}}]{Orth:2005kq}
\bibinfo{author}{\bibfnamefont{B.}~\bibnamefont{Orth}},
  \bibinfo{author}{\bibfnamefont{T.}~\bibnamefont{Lippert}}, \bibnamefont{and}
  \bibinfo{author}{\bibfnamefont{K.}~\bibnamefont{Schilling}},
  \bibinfo{journal}{Phys.Rev.} \textbf{\bibinfo{volume}{D72}},
  \bibinfo{pages}{014503} (\bibinfo{year}{2005}), \eprint{hep-lat/0503016}.

\bibitem[{\citenamefont{Sommer}(1994)}]{Sommer:1993ce}
\bibinfo{author}{\bibfnamefont{R.}~\bibnamefont{Sommer}},
  \bibinfo{journal}{Nucl.Phys.} \textbf{\bibinfo{volume}{B411}},
  \bibinfo{pages}{839} (\bibinfo{year}{1994}), \eprint{hep-lat/9310022}.

\bibitem[{\citenamefont{Clark et~al.}(2010)\citenamefont{Clark, Babich, Barros,
  Brower, and Rebbi}}]{Clark:2009wm}
\bibinfo{author}{\bibfnamefont{M.}~\bibnamefont{Clark}},
  \bibinfo{author}{\bibfnamefont{R.}~\bibnamefont{Babich}},
  \bibinfo{author}{\bibfnamefont{K.}~\bibnamefont{Barros}},
  \bibinfo{author}{\bibfnamefont{R.}~\bibnamefont{Brower}}, \bibnamefont{and}
  \bibinfo{author}{\bibfnamefont{C.}~\bibnamefont{Rebbi}},
  \bibinfo{journal}{Comput.Phys.Commun.} \textbf{\bibinfo{volume}{181}},
  \bibinfo{pages}{1517} (\bibinfo{year}{2010}), \eprint{0911.3191}.

\bibitem[{\citenamefont{Alexandrou et~al.}(1994)\citenamefont{Alexandrou,
  Gusken, Jegerlehner, Schilling, and Sommer}}]{Alexandrou:1992ti}
\bibinfo{author}{\bibfnamefont{C.}~\bibnamefont{Alexandrou}},
  \bibinfo{author}{\bibfnamefont{S.}~\bibnamefont{Gusken}},
  \bibinfo{author}{\bibfnamefont{F.}~\bibnamefont{Jegerlehner}},
  \bibinfo{author}{\bibfnamefont{K.}~\bibnamefont{Schilling}},
  \bibnamefont{and} \bibinfo{author}{\bibfnamefont{R.}~\bibnamefont{Sommer}},
  \bibinfo{journal}{Nucl. Phys.} \textbf{\bibinfo{volume}{B414}},
  \bibinfo{pages}{815} (\bibinfo{year}{1994}), \eprint{hep-lat/9211042}.

\bibitem[{\citenamefont{Gusken}(1990)}]{Gusken:1989}
\bibinfo{author}{\bibfnamefont{S.}~\bibnamefont{Gusken}},
  \bibinfo{journal}{Nucl. Phys. Proc. Suppl.} \textbf{\bibinfo{volume}{17}},
  \bibinfo{pages}{361} (\bibinfo{year}{1990}).

\bibitem[{\citenamefont{Alexandrou et~al.}(2008)}]{Alexandrou:2008tn}
\bibinfo{author}{\bibfnamefont{C.}~\bibnamefont{Alexandrou}}
  \bibnamefont{et~al.} (\bibinfo{collaboration}{European Twisted Mass
  Collaboration}), \bibinfo{journal}{Phys. Rev.}
  \textbf{\bibinfo{volume}{D78}}, \bibinfo{pages}{014509}
  (\bibinfo{year}{2008}), \eprint{0803.3190}.

\bibitem[{\citenamefont{Struckmann et~al.}(2001)}]{Struckmann:2000bt}
\bibinfo{author}{\bibfnamefont{T.}~\bibnamefont{Struckmann}}
  \bibnamefont{et~al.} (\bibinfo{collaboration}{TXL Collaboration, T(X)L
  Collaboration}), \bibinfo{journal}{Phys.Rev.} \textbf{\bibinfo{volume}{D63}},
  \bibinfo{pages}{074503} (\bibinfo{year}{2001}), \eprint{hep-lat/0010005}.

\bibitem[{\citenamefont{Becirevic et~al.}(2006)\citenamefont{Becirevic,
  Blossier, Boucaud, Gimenez, Lubicz et~al.}}]{Becirevic:2005ta}
\bibinfo{author}{\bibfnamefont{D.}~\bibnamefont{Becirevic}},
  \bibinfo{author}{\bibfnamefont{B.}~\bibnamefont{Blossier}},
  \bibinfo{author}{\bibfnamefont{P.}~\bibnamefont{Boucaud}},
  \bibinfo{author}{\bibfnamefont{V.}~\bibnamefont{Gimenez}},
  \bibinfo{author}{\bibfnamefont{V.}~\bibnamefont{Lubicz}},
  \bibnamefont{et~al.}, \bibinfo{journal}{Nucl.Phys.}
  \textbf{\bibinfo{volume}{B734}}, \bibinfo{pages}{138} (\bibinfo{year}{2006}),
  \eprint{hep-lat/0510014}.

\bibitem[{\citenamefont{Alexandrou et~al.}(2006)\citenamefont{Alexandrou,
  Koutsou, Negele, and Tsapalis}}]{Alexandrou:2006ru}
\bibinfo{author}{\bibfnamefont{C.}~\bibnamefont{Alexandrou}},
  \bibinfo{author}{\bibfnamefont{G.}~\bibnamefont{Koutsou}},
  \bibinfo{author}{\bibfnamefont{J.~W.} \bibnamefont{Negele}},
  \bibnamefont{and} \bibinfo{author}{\bibfnamefont{A.}~\bibnamefont{Tsapalis}},
  \bibinfo{journal}{Phys. Rev.} \textbf{\bibinfo{volume}{D74}},
  \bibinfo{pages}{034508} (\bibinfo{year}{2006}), \eprint{hep-lat/0605017}.

\bibitem[{\citenamefont{Giedt et~al.}(2009)\citenamefont{Giedt, Thomas, and
  Young}}]{Giedt:2009mr}
\bibinfo{author}{\bibfnamefont{J.}~\bibnamefont{Giedt}},
  \bibinfo{author}{\bibfnamefont{A.~W.} \bibnamefont{Thomas}},
  \bibnamefont{and} \bibinfo{author}{\bibfnamefont{R.~D.} \bibnamefont{Young}},
  \bibinfo{journal}{Phys.Rev.Lett.} \textbf{\bibinfo{volume}{103}},
  \bibinfo{pages}{201802} (\bibinfo{year}{2009}), \eprint{0907.4177}.

\bibitem[{\citenamefont{Koch}(1982)}]{Koch:1982pu}
\bibinfo{author}{\bibfnamefont{R.}~\bibnamefont{Koch}},
  \bibinfo{journal}{Z.Phys.} \textbf{\bibinfo{volume}{C15}},
  \bibinfo{pages}{161} (\bibinfo{year}{1982}).

\bibitem[{\citenamefont{Pavan et~al.}(2002)\citenamefont{Pavan, Strakovsky,
  Workman, and Arndt}}]{Pavan:2001wz}
\bibinfo{author}{\bibfnamefont{M.}~\bibnamefont{Pavan}},
  \bibinfo{author}{\bibfnamefont{I.}~\bibnamefont{Strakovsky}},
  \bibinfo{author}{\bibfnamefont{R.}~\bibnamefont{Workman}}, \bibnamefont{and}
  \bibinfo{author}{\bibfnamefont{R.}~\bibnamefont{Arndt}},
  \bibinfo{journal}{PiN Newslett.} \textbf{\bibinfo{volume}{16}},
  \bibinfo{pages}{110} (\bibinfo{year}{2002}), \eprint{hep-ph/0111066}.

\bibitem[{\citenamefont{Gusken et~al.}(1999)}]{Gusken:1998wy}
\bibinfo{author}{\bibfnamefont{S.}~\bibnamefont{Gusken}} \bibnamefont{et~al.}
  (\bibinfo{collaboration}{TXL Collaboration}), \bibinfo{journal}{Phys.Rev.}
  \textbf{\bibinfo{volume}{D59}}, \bibinfo{pages}{054504}
  (\bibinfo{year}{1999}), \eprint{hep-lat/9809066}.

\bibitem[{\citenamefont{Alexandrou et~al.}(2009)}]{Alexandrou:2009qu}
\bibinfo{author}{\bibfnamefont{C.}~\bibnamefont{Alexandrou}}
  \bibnamefont{et~al.} (\bibinfo{collaboration}{ETM Collaboration}),
  \bibinfo{journal}{Phys. Rev.} \textbf{\bibinfo{volume}{D80}},
  \bibinfo{pages}{114503} (\bibinfo{year}{2009}), \eprint{0910.2419}.

\bibitem[{\citenamefont{Young and Thomas}(2010)}]{Young:2009zb}
\bibinfo{author}{\bibfnamefont{R.}~\bibnamefont{Young}} \bibnamefont{and}
  \bibinfo{author}{\bibfnamefont{A.}~\bibnamefont{Thomas}},
  \bibinfo{journal}{Phys.Rev.} \textbf{\bibinfo{volume}{D81}},
  \bibinfo{pages}{014503} (\bibinfo{year}{2010}), \eprint{0901.3310}.

\bibitem[{\citenamefont{Walker-Loud et~al.}(2009)}]{WalkerLoud:2008bp}
\bibinfo{author}{\bibfnamefont{A.}~\bibnamefont{Walker-Loud}}
  \bibnamefont{et~al.}, \bibinfo{journal}{Phys.Rev. D}
  \textbf{\bibinfo{volume}{79}} (\bibinfo{year}{2009}), \eprint{0806.4549}.

\bibitem[{\citenamefont{Ohki et~al.}(2008)\citenamefont{Ohki, Fukaya,
  Hashimoto, Kaneko, Matsufuru et~al.}}]{Ohki:2008ff}
\bibinfo{author}{\bibfnamefont{H.}~\bibnamefont{Ohki}},
  \bibinfo{author}{\bibfnamefont{H.}~\bibnamefont{Fukaya}},
  \bibinfo{author}{\bibfnamefont{S.}~\bibnamefont{Hashimoto}},
  \bibinfo{author}{\bibfnamefont{T.}~\bibnamefont{Kaneko}},
  \bibinfo{author}{\bibfnamefont{H.}~\bibnamefont{Matsufuru}},
  \bibnamefont{et~al.}, \bibinfo{journal}{Phys.Rev.}
  \textbf{\bibinfo{volume}{D78}}, \bibinfo{pages}{054502}
  (\bibinfo{year}{2008}), \eprint{0806.4744}.

\bibitem[{\citenamefont{Aoki et~al.}(2009)}]{Aoki:2008sm}
\bibinfo{author}{\bibfnamefont{S.}~\bibnamefont{Aoki}} \bibnamefont{et~al.}
  (\bibinfo{collaboration}{PACS-CS}), \bibinfo{journal}{Phys. Rev. D}
  \textbf{\bibinfo{volume}{79}}, \bibinfo{pages}{034503}
  (\bibinfo{year}{2009}), \eprint{0807.1661}.

\bibitem[{\citenamefont{Durr et~al.}(2010)\citenamefont{Durr, Fodor, Frison,
  Hemmert, Hoelbling et~al.}}]{Durr:2010ni}
\bibinfo{author}{\bibfnamefont{S.}~\bibnamefont{Durr}},
  \bibinfo{author}{\bibfnamefont{Z.}~\bibnamefont{Fodor}},
  \bibinfo{author}{\bibfnamefont{J.}~\bibnamefont{Frison}},
  \bibinfo{author}{\bibfnamefont{T.}~\bibnamefont{Hemmert}},
  \bibinfo{author}{\bibfnamefont{C.}~\bibnamefont{Hoelbling}},
  \bibnamefont{et~al.}, \bibinfo{journal}{PoS}
  \textbf{\bibinfo{volume}{LATTICE2010}}, \bibinfo{pages}{102}
  (\bibinfo{year}{2010}), \eprint{1012.1208}.

\bibitem[{\citenamefont{Evans et~al.}(2010)\citenamefont{Evans, Bali, and
  Collins}}]{Evans:2010tg}
\bibinfo{author}{\bibfnamefont{R.}~\bibnamefont{Evans}},
  \bibinfo{author}{\bibfnamefont{G.}~\bibnamefont{Bali}}, \bibnamefont{and}
  \bibinfo{author}{\bibfnamefont{S.}~\bibnamefont{Collins}},
  \bibinfo{journal}{Phys.Rev.} \textbf{\bibinfo{volume}{D82}},
  \bibinfo{pages}{094501} (\bibinfo{year}{2010}), \eprint{1008.3293}.

\end{thebibliography}

\end{document}